\documentclass[a4paper,fleqn]{cas-dc}

\usepackage[authoryear]{natbib}

\usepackage{pifont}
\usepackage{algpseudocode}
\usepackage{algorithmicx,algorithm}
\usepackage{subfigure}
\def\tsc#1{\csdef{#1}{\textsc{\lowercase{#1}}\xspace}}
\tsc{WGM}
\tsc{QE}
\tsc{EP}
\tsc{PMS}
\tsc{BEC}
\tsc{DE}

\begin{document}
\let\WriteBookmarks\relax
\def\floatpagepagefraction{1}
\def\textpagefraction{.001}



\title [mode = title]{Optimal Control of Malware Propagation in IoT Networks}                      


%
\author[1]{Jafar}[auid=000,bioid=1,
                        prefix=Mousa Tayseer,
                        orcid=0000-0002-0408-0541]

\cormark[1]


\ead{m.tayseermousajafar@deakin.edu.au}



\affiliation[1]{organization={The School of Information Technology, Deakin University},
    city={Melbourne},
    postcode={VIC,3125}, 
    country={Australia}}

\author[1]{Yang}[prefix=Lu-Xing,
                        orcid=0000-0002-9229-5787]

\ead{y.luxing@deakin.edu.au}

\author[1]{Li}[prefix=Gang,
                        orcid=0000-0003-1583-641X]
\ead{gang.li@deakin.edu.au}

\credit{Data curation, Writing - Original draft preparation}

\affiliation[2]{organization={The School of Big Data and Software Engineering, Chongqing University},
    city={Chongqing},
    postcode={400044}, 
    state={P.R.},
    country={China}}

\author[2]{Yang}[prefix=Xiaofan,
                        orcid=0000-0001-6931-2692]
\ead{xfyang1964@cqu.edu.cn}

\cortext[cor1]{Corresponding author}


\begin{abstract}
The rapid proliferation of Internet of Things (IoT) devices in recent years has resulted in a significant surge in the number of cyber-attacks targeting these devices. Recent data indicates that the number of such attacks has increased by over 100 percent, highlighting the urgent need for robust cybersecurity measures to mitigate these threats. 
In addition, a cyber-attack will begin to spread malware across the network once it has successfully compromised an IoT network. However, to mitigate this attack, a new patch must be applied immediately. In reality, the time required to prepare and apply the new patch can vary significantly depending on the nature of the cyber-attack. In this paper, we address the issue of how to mitigate cyber-attacks before the new patch is applied by formulating an optimal control strategy that reduces the impact of malware propagation and minimise the number of infected devices across IoT networks in the smart home. A novel node-based epidemiological model susceptible, infected high, infected low, recover first, and recover complete $ \mathbf{(SI_HI_LR_FR_C)}$ is established with immediate response state for the restricted environment. After that, the impact of malware on IoT devices using both high and low infected rates will be analyzed. Finally, to illustrate the main results, several numerical analyses are carried out in addition to simulate the real-world scenario of IoT networks in the smart home, we built a dataset to be used in the experiments.
\end{abstract}




\maketitle

\section{Introduction}

Malware continues to pose a significant and pervasive threat to cybersecurity, representing one of the most pressing issues facing the field today. The sheer volume of malware has surged in recent years, with a staggering (1,069,281,538) instances recorded as of August 1st, 2023 \cite{avatlasAVATLASMalware}. This alarming trend underscores the urgent need for heightened vigilance and robust strategies to combat the spread of malware, particularly in the context of IoT networks. 

IoT represents the most significant digital mega-trend that bridges the physical and virtual worlds. Over the past few years, the deployment of IoT devices has experienced an exponential growth rate, unlike any other technological advancement in history. By 2025, it is projected that the number of IoT devices connected to the internet will exceed 75 billion \cite{jamshed2022challenges}. IoT networks have become an essential component of our daily lives, from smart cities and homes to healthcare systems. These networks connect numerous smart devices, allowing them to gather and share data, and they are present in various locations  \cite{ray2018survey}. However, cybercriminals have started to target IoT networks more frequently, making them an increasingly attractive target for cyber-attacks.

Entire IoT devices are vulnerable to being infected with malware to launch cybersecurity attacks. For instance the Mirai botnet, a global epidemic that happened in 2016, has targeted over 600,000 IoT devices to establish an enormous distributed denial-of-service (DDoS) attack \cite{xenofontos2021consumer}. In the aftermath of this infamous incident, numerous malware families have evolved into existence to carry out more sophisticated attacks. In the second half of 2022 another incident has taken place, hackers have exploited a security vulnerability in the Realtek Jungle SDK to perform a critical remote code execution, where 134 million attacks attempting to infect smart devices \cite{chatzoglou2022your}. This vulnerability was exploited by a new type of malware known as RedGoBot.

\subsection{Motivation and problem statement}
In recent years, there has been a surge in cyber-attacks targeting IoT devices, with an increase of over 100\% \cite{ashraf2022survey}. Cybercriminals have turned their attention to this area and have developed sophisticated methods to breach IoT networks. Once a network is compromised, malware can quickly spread throughout the system, necessitating the installation of a new patch to mitigate the attack. The significant challenge in combating malware is the time, and cost required to develop an effective vaccine or patch. Developing an effective vaccine or patch for IoT malware can take a long time. In addition to the high cost of preparing a new patch, which depends on the type of malware and the type of cybersecurity attack. However, the process of preparing and installing the patch can be time-consuming, thereby exacerbating the risk posed by the attack.

The underlying problem that needs to be addressed is the identification of an optimal control solution that incorporates an immediate response mechanism to counteract the propagation of malware over IoT networks. It is imperative to develop an effective approach that can effectively mitigate the impact of such malware propagation across these networks. To the best of our knowledge, there has been no previous work that has specifically addressed this problem.

This paper aims to achieve this objective by introducing a novel node-based epidemiological model, called susceptible, infected high, infected low, recover first, and recover complete $\mathbf{(SI_HI_LR_FR_C)}$ model, which incorporates an immediate response state for the restricted environment. 
The recovery first state is designed to provide an immediate response to cybersecurity incidents and to serve as a viable defense strategy against early-stage malware attacks on the IoT network. This approach aims to effectively mitigate the impact of cyber threats at an early stage, and thus prevent further damage to the network.
On the other hand, the recovery complete state is achieved by applying the latest and most powerful cybersecurity patch to address cybersecurity incidents. This approach ensures that all infected IoT devices receive the most up-to-date patch, which enhances their overall security posture.

Furthermore, the key limitation of existing epidemic models is the assumption that infected devices have the same infection rate or ability to spread malware over the network. However, in reality, these rates may vary depending on the specific characteristics of the malware and its mode of propagation, underscoring the need for more nuanced modeling approaches to accurately capture the dynamics of malware spreading. 
We will address this issue by classifying the IoT devices into two states, the first state is a high infected state, which will assist to determine the IoT devices that have strong spreading capability. The second state is a low infected state, which will assist to determine the IoT devices that have weak spreading capability. Thus will have a positive impact on the mitigation of the propagation process through the reduction of the spread of malware.

\subsection{Contributions}

The main contributions that were made in this paper are outlined below:
\begin{itemize}
\item Firstly, the proposed model introduces a novel idea, the recovery first state (immediate response), which will enhance the mitigation of the propagation process by controlling the malware propagation before the new patch is applied. This strategic approach is designed to increase the effectiveness of containment measures.  

\item Secondly, the proposed model classifies the IoT devices into two distinct states, where the first state represents a high infected state and encompasses IoT devices exhibiting strong spreading capability. In contrast, the second state represents a low infected state and includes IoT devices with weak spreading capability.

\item Finally, the optimal control problem has been designed based on Pontryagin\textquotesingle s Minimisation Principle, with the aim of minimizing the spread of malware and reducing the associated costs of patching across IoT networks. By formulating the problem in this way, the proposed approach seeks to identify the optimal control strategy that balances the competing objectives of malware containment and cost efficiency. This approach can enable more effective and efficient management of cybersecurity threats in the rapidly evolving landscape of IoT devices and networks.

\end{itemize}

The structure of the paper can be summarised as follows: \autoref{Related} provides related work, which includes an overview of epidemiological models of malware propagation and an optimal control-theoretic approach to mitigating the spread of malware. \autoref{Formation} describes the formation of a node-level $ \mathbf{SI_HI_LR_FR_C}$ model. \autoref{Optimal} outlines an optimal control strategy that demonstrates the existence of an optimal control and the optimality of the system. \autoref{Experiments} presents the results of experiments and their analysis, followed by a conclusion in \autoref{Conclusion}.

\section{Related work}
\label{Related}

The purpose of this section is to provide an overview of the work that has been done in relation to our work from two perspectives, firstly from epidemiological models of malware propagation, and secondly from an optimal control theoretic approach to mitigating the spread of malware.

\subsection{ Epidemiology models for malware propagation}

The first appearance of the epidemiological model was in the third quarter of the seventeenth century, where the model was used to study the spread of smallpox \cite{bernoulli1760essai}. Researchers have continued to introduce new models to study and analyse the spread process in many different fields, including biology, engineering, mathematics, social sciences, economics, physics and computer science. 
 
The susceptible, exposed, infected, recovered, dead \newline(SEIRD) model was introduced in {\cite{chen2022mobility}}, the model analyses the dynamics of IoT malware propagation through various communication channels. They performed mathematical calculations to determine the threshold for malware transmission, which was used as a benchmark for malware propagation. The authors relied on the opening rate of the malicious file to spread the malware, and also used an imprecise mechanism to determine the spread of the malware as a threshold transmission. 

In { \cite{xia2020dynamic}}, the authors proposed ignorant, dks,hn-exposed, pvirus, spread, recover (IDEPSR) model to study virus propagation in the smart city based on the intelligent devices, identification and propagation capabilities were used as social attributes. Furthermore, they assumed two types of contributions to determine the propagation ability of the infected node, direct and indirect for the neighbour contribution, thus adding more complexity. 

In  \cite{xia2020modeling} , the IoT-BSI model was built based on ignorant, exposed, spreader, recover (IESR) states to investigate the identification ability and propagation capability in the social IoT for botnet propagation. The model has been applied to smart devices by assuming that the devices will not be reinfected while only one case has been taken into account: once the smart devices have been cured, they cannot be re-infected.

The authors in \cite{yan2021modeling}  studied malware spreading over the wireless IoT devices by proposed a new epidemic model heterogeneous susceptible, exposed, infected, recovered (HSEIR). The threshold was determined based on the basic reproduction number for malware spread or not.They also used a fuzzy method to determine whether infected devices were capable of transmitting, which was less accurate.

The susceptible, exposed, infective, recovered (SEIRS) model was used in  \cite{miao2021stochastic} for edge computing based IoT to study the spread of the malware by considering the stochastic characteristic. The non-cooperative, susceptible, infectious, dead (NSID) epidemic model was established in  \cite{zhang2019preventing} to illustrate the propagation of malware with strategic mobile users over the device-to-device (D2D) network. From the perspective of a secure node, the authors examined the isolation process and considered the attacker\textquotesingle s strategy of killing infected devices to reduce the spread of malware. Where the strategy of the attack is assumed to be fixed and acknowledged.   

The susceptible, infectious, cured, dead (SICD) epidemic model was extended in  \cite{zhang2019differential} by adding two non-cooperative states for infectious and cured states to study malware propagation for mobile networks under the risk of epidemic. However, their model suppose that the new device will switch to the non-cooperative level in order to reduce the number of infected devices.   

The susceptible, exposed, infectious, recovered (SEIR) model explored in  \cite{zhang2022hopf} to investigate mobile wireless sensor networks by conducting an in-depth analysis of malware propagation. The evaluation of the fundamental reproduction number of the model is accomplished through the utilization of the next generation matrix method.

In  \cite{muthukrishnan2021optimal}, the spread of malware in wireless sensor networks was investigated using the susceptible, infected, traced, patched, susceptible (SITPS) epidemic model. Based on a node level, the model has been used to study the tracking and patching of malware. Moreover, the range of the communication was used as a point of analysis for the spread malware over wireless sensor networks.

The authors in  \cite{dong2022dynamical} introduced a new model with two exposed states, exposed type- 1, and exposed type-2 in the susceptible, exposed-1, exposed-2, infectious, quarantine, recovered (SE1E2IQR) epidemic model to investigate the heterogeneity and the effectiveness of quarantine strategy for malware propagation in wireless sensor network. Furthermore, the analysis and evaluation were simplified by assuming a fixed number of sensors and the same rate of disease between nodes.

Researchers have also developed epidemiological models based on the classic susceptible, infected, susceptible (SIS) and susceptible, infected, recovered (SIR) models with a fixed number of states to control the spread of malware. A comparison of attributes between the proposed model and existing models is presented in 
\autoref{tab:Table1}.
However, the limitation of the SIR model is that the number of states is insufficient to describe the behaviour of malware in the IoT network \cite{7393962}. To the best of our knowledge, none of the existing models have applied an immediate response state to the malware propagation over IoT networks, which will have a significant impact on the process of malware propagation. In this paper, the proposed model will address this issue to improve the propagation of malware in the IoT networks.

\begin{table} [!h]
\centering
\caption{Comparative of the attributes between the proposed model and existing models.}
\centering
\begin{tabular}{ >{\centering\arraybackslash} m{1.5cm} m{2.7 cm}>{\centering\arraybackslash}m{1.3 cm}>{\centering\arraybackslash}m{1.3 cm}  }
\hline
Ref &  Model & Immediate Response & Classifying high and low infected \\ \hline 
 \cite{chen2022mobility} & Susceptible,Exposed, Infected,Recovered, Dead(SEIRD)& \ding{55}  & \ding{55}\\ \hline
 \cite{zhang2022hopf}& Susceptible,Exposed, Infectious, Recovered (SEIR)  &  \ding{55} &    \ding{55}\\ \hline
 \cite{miao2021stochastic} & Susceptible,Exposed, Infective,Recovered (SEIRS)  & \ding{55} &  \ding{55}  \\ \hline
\cite{yan2021modeling} & Heterogeneous, Susceptible, Exposed,Infected, Recovered (HSEIR) &  \ding{55} &  \ding{55} \\ \hline
 \cite{muthukrishnan2021optimal} & Susceptible,Infected, Traced,Patched, Susceptible (SITPS) &  \ding{55} &  \ding{55} \\ \hline
 \cite{xia2020modeling} & Ignorant,Exposed, Spreader,Recover  (IESR)  &  \ding{55}&  \ding{55} \\ \hline
{ \cite{xia2020dynamic}}  & Ignorant,DKsHN, Exposed,Pvirus, Spread,Recover (IDEPSR)  & \ding{55} &  \ding{55}  \\ \hline
 \cite{zhang2019preventing} & Non-cooperative, Susceptible, Infectious,Dead (NSID)     & \ding{55} &  \ding{55} \\ \hline
 \cite{zhang2019differential} & Susceptible,Infectious, Cured,Dead (SICD) & \ding{55} & \ding{55}  \\ \hline
 \cite{dong2022dynamical} & Susceptible,Exposed1,  Exposed2, Infectious, Quarantine, Recovered (SE1E2IQR)  &  \ding{55} & \ding{55}\\ \hline
Proposed model & Susceptible,Infected-High,Infected-Low,  Recover-First, Recover-Complete $ \mathbf{(SI_HI_LR_FR_C)}$ &    \checkmark  & \checkmark \\ \hline
\end{tabular}
\label{tab:Table1}
\vspace*{-3mm}
\end{table}

\subsection{Optimal control theoretic approach to mitigate the spread of malware}
Optimal control theory is considered as a one of the mathematical optimization branch, which is concerned with finding out how to control a dynamic system over a period of time by minimising or maximising the objective functions \cite{rao2009survey}. Existing optimal control theoretical models for mitigating the spread of malware can be classified into three categories:
1) optimal control models at the population level,  2) optimal control models at the network level, and 3) optimal control models at the individual level. These models are briefly described below.

In an optimal control model at the population level, the classification of individuals is determined by their particular states, whereas the size of each group experiences a unique evolution that is governed by a separate equation.
The authors in {\cite{liu2021dynamical}} investigated the local and global stability of the system by building the (SEIR) propagation model for the spreading virus in wireless sensor networks. They constructed the Hamiltonian function based on the Pontryagin Maximum Principle to find the optimal control strategy that minimises the cost function. Furthermore, a comprehensive survey at the population level models presented in {\cite{7393962}; \cite{zino2021analysis}; {\cite{huang2022game}.

In an optimal control model at the network level: based on the type of network topology, such as complex network and scale-free network, the optimal control strategy is formulated. The optimal control problem was formulated on the (SIVRS) model that was designed for the spread of virus variation, to maximise the recovered state over the complex network by applying Pontryagin\textquotesingle s Minimum Principle in {\cite{xu2017optimal}. Another work targeting virus propagation over the complex network was presented by the authors in {\cite{zhang2018global}, where they investigated the impact of combining the network topology with countermeasures. The optimal control problem was analysed to find the optimal control that minimises the density of the infected nodes, and the total budget for the countermeasures.
For scale-free networks,  the authors in \cite{zhang2016optimal} have been studied the optimal control strategy for the computer virus in the scale-free networks. They analysed the global stability and impact of network topology with reinstallation system through the construction of a new (SLBOS) model of virus. 

Node-level propagation models are designed to capture the dynamics of information propagation within a network, treating each individual node as a distinct class. These models aim to provide valuable insights into the evolutionary trends of the expected network state, allowing for a comprehensive analysis of information diffusion patterns at the node level \cite{yu2021modeling, overton2022approximating, bi2022defense}. A noteworthy advantage of node-level propagation models is their ability to effectively capture and characterize propagation phenomena that occur in diverse network structures, providing accurate and comprehensive insights into the dynamics of such phenomena. As wireless and mobile communication technologies continue to advance, many IoT networks exhibit irregular topologies, which pose unique challenges and complexities in their design and operation \cite{vaezi2022cellular}.

In an optimal control model at the node level: the optimal control problem to discover the best balance between the cost of developing a new patch and the impact of the virus for computer viruses has been studied in  \cite{huang2019seeking}  based on the nodel level. Another work presented in  \cite{yang2016optimal}, the authors formulated an optimal control problem to capture the optimal dynamic immunisation using a node-level SIRS model.

The issue of optimal control strategies for IoT networks has been relatively overlooked in the current literature. Based on our thorough review of the existing literature and to the best of our knowledge, there is limited research on applying of optimal control strategies for IoT networks. Therefore, this paper aims to address this gap by formulating a node-level optimal control problem that aims to minimise the number of infected IoT devices, while reducing the overall costs associated with implementing both the restricted environment and the latest patch.

\section{Formation of a node-level $ \mathbf{SI_HI_LR_FR_C}$ model }
\label{Formation}
In this section, we present the formulation of the epidemiological node-level $ \mathbf{SI_HI_LR_FR_C}$ model and elucidate the probabilistic assumptions that underpin the model, outlining the transition dynamics within the IoT network.

We consider IoT network with a population of $N $ IoT devices labelled $1, 2, ..., N.$ That is represented in the graph $G = (V, E)$ which is denote the topology of IoT network, where each node stands for an IoT device $V={1,2,...,N}$ and $(i,j) \in E $ if and only if IoT device $i$ has connected to IoT device $j$. Its adjacency matrix $A= [a_{ij}]_{N*N} $ can describe the graph $G$ for IoT network, where $a_{ij}= 0 $ or $1 $ depending on $(i,j) \in E $ or not. The parameters of the $ \mathbf{SI_HI_LR_FR_C}$ model that are used in the propagation processes are described in  \autoref{tab:Table2}.

The proposed $ \mathbf{SI_HI_LR_FR_C}$ model assume that the population of IoT devices infected by the IoT device $j$ where $ \beta_{L}$ is a weak spreading capability whereas the population of IoT devices infected by the IoT device $j$ where $ \beta_{H}$ is a strong spreading capability where $0\leq \beta_{L} \leq \beta_{H}$.
At any time every IoT device in the population $N$ is in one of five possible states: susceptible, infected high, infected low, recover first , and recover complete.

\textbf{Susceptible} $\mathbf{(S_i(t))}$ : all IoT devices are susceptible and have not been infected by malware but can be attacked by malware.
\textbf{Infected High} $\mathbf{(I_{H_i}(t))}$: IoT device was infected with malware and the device has fast spreading capability to spread the malware over IoT network.
\textbf{Infected Low} $\mathbf{(I_{L_i}(t))}$: IoT device was infected with malware and the device has low spreading capability to spread the malware over IoT network.
\textbf{Recover First}
$\mathbf{(R_{F_i}(t))}$: IoT device respond immediately to cybersecurity incidents by applying restricted environment concept and other security measures.
\textbf{Recover Complete} $\mathbf{(R_{C_i}(t))}$: security patch has been applied and provides the technique for IoT devices after malware removal from IoT network.

The state of the IoT network at time $t$ can be probabilistically captured by the vector:
\begin{equation}
       \begin{split}
            & (S_1(t),…,S_N(t),I_{H_1}(t),…,I_{H_N}(t),I_{L_1}(t),…,  \\
            & I_{L_N}(t),R_{F_1}(t),…,R_{F_N}(t),R_{C_1}(t),…,R_{C_N}(t))^T. \\
        \end{split} 
\end{equation}
For simplicity, state (1) can be expressed as follows:
\begin{equation}
      \begin{split}
             & \mathbf{S}(t)=(S_1(t),…,S_N(t))^T \; , \\
             & \mathbf{I_H}(t)=(I_{H_1}(t),…,I_{H_N}(t))^T \; ,  \\
             & \mathbf{I_L}(t)=(I_{L_1}(t),…,I_{L_N}(t))^T  \; , \\
             &  \mathbf{R_F}(t)=(R_{F_1}(t),…,R_{F_N}(t))^T \; ,  \\ 
             &  \mathbf{R_C}(t)=(R_{C_1}(t),…,R_{C_N}(t))^T \;  .\\ 
       \end{split} 
\end{equation}
In order to achieve normalization, it is imperative that the densities of nodes in all five states adhere to the normalization requirement.
\begin{equation}
        S_i(t)+I_{H_i}(t)+I_{L_i}(t)+R_{F_i}(t)+R_{C_i}(t)=1  \;\; ,1\leq i \leq N .
\end{equation}

\begin{table} [!h]
\centering
\caption{List of model parameters that are used in the propagation processes.}
\centering
\begin{tabular}{>{\centering\arraybackslash} m{1cm}>{\centering\arraybackslash} m{1.2 cm} m{4.80cm}  }\hline
&  &  \\
S. No & Parameters & Description of the parameters\\    
&  &  \\ \hline
1. &   $ N    $       & Total number of IoT devices at the time $t$ in the IoT network. \\ \hline
2. &   $S_i(t)  $     & Probability of susceptible IoT devices at the time $t$ in the IoT network. \\ \hline
3. &  $I_{H_i}(t)$    & Probability of infected high IoT devices at the time $t$ in the IoT network. \\ \hline
4. &   $I_{L_i}(t)$   & Probability of infected low IoT devices at the time $t$ in the IoT network. \\ \hline
5. &   $R_{F_i}(t)$   & Probability of secured IoT devices (first response) at the time $t$ in the IoT network. \\ \hline
6. &   $R_{C_i}(t)$   & Probability of secured IoT devices (recovery complete) at the time $t$ in the IoT network. \\ \hline
7. &   $\beta_{H}$    & Infection rate of IoT devices with weak spreading capabilities in the IoT network. \\ \hline
8. &   $\beta_{L}$    & Infection rate of IoT devices with strong spreading capabilities in the IoT network. \\ \hline
9. &  $\delta_{i}$    & Probability that recovery first IoT devices may be become recovery complete after applying latest cybersecurity patch at the time $t$ in the IoT network. \\ \hline
10. &  $\gamma_{H_i}$  & Probability that infected high IoT devices may be become recovery first IoT devices after applying the security measures (prevent broadcast and multicast traffic) at the time $t$ in the IoT network. \\ \hline
11. &  $\gamma_{L_i}$ & Probability that infected low IoT devices may be become recovery first IoT devices after applying the security measures (prevent broadcast traffic) at the time $t$ in the IoT network. \\ \hline
12. &  $G $       & The graph for the topology of the IoT networks. \\ \hline
13. &  $A$      & Adjacency matrix to describe graph $G$ in IoT networks $[a_{ij}]_{N*N}$.   \\ \hline
&  &  \\
14. &  $ [0, T] $       & The interval time in the population $N$. \\ 
&  &  \\ \hline
15. &  $ \mathbf{u}$ & The control function $ \mathbf{u}(t) \in \mathbf{U}$ where $ \mathbf{u} = (\delta_1,...,\delta_N,  \gamma_{H_1},...,\gamma_{H_N}, \gamma_{L_1},...,\gamma_{L_N})^T  $. \\ \hline
\end{tabular}
\label{tab:Table2}
\vspace*{-3mm}
\end{table}

The probabilistic assumptions for our model that explain the transition process between IoT devices are presented as following:

{\textbf{ (A1) Susceptible to Infected Low}} $\mathbf{(S_i(t)\rightarrow I_{L_i}(t))}$:
 at any time, due to the cybersecurity attack that contain malware with weak spreading capability, susceptible IoT device $i$ transit into infected low stage by an infected device $j$ when the malware actively at an average rate: 
\begin{equation}
\label{deqn_ex1a}
 \beta_{L}\; \sum_{j=1}^{N} a_{ij} \; I_{L_j}(t),  \qquad 1\leq i \leq N ,\; 0\leq \beta_{L} \leq \beta_{H}, 
\end{equation}
where $\beta_{L}$is malware infection rate of IoT devices with weak spreading capability.}

{\textbf{(A2) Susceptible to Infected High}} $\mathbf{(S_i(t) \rightarrow I_{H_i}(t))}$:
at any time, due to the cybersecurity attack that contain malware with strong spreading capability, susceptible IoT device $i$ transit into infected high stage by an infected device $j$ when the malware actively at an average rate:
\begin{equation}
\label{deqn_ex1a}
\beta_{H} \; \sum_{j=1}^{N} a_{ij} \;  I_{H_j}(t),  \qquad 1\leq i \leq N , \; \beta_{H} \geq 0 , 
\end{equation}
where $\beta_{H}$is malware infection rate of IoT devices with strong spreading capability.

{\textbf{(A3) Infected High to Recovery First}} $\mathbf{(I_{H_i}(t) \rightarrow R_{F_i}(t))}$: 
 at any time, due to immediately respond to cybersecurity incidents and some sufficient defence mechanisms that used to deal with malware at the early stage in IoT network, the IoT device infected high transit into recovery first stage at an average rate:
\begin{equation}
\label{deqn_ex1a}
\gamma_{H_i}(t),   \qquad 1\leq i \leq N, 
\end{equation}
where $\gamma_{H_i}$ is the rate of applying the security measures that assist the compromised IoT devices to reduce the impact of the cybersecurity attack over IoT network. (Restricted environment by prevent broadcast and multicast traffic for the infected devices and other security defence mechanisms). \newline
$\gamma_{H_i}(t) \in L^2[0, T]$, The symbol $L^2[0, T]$ stands for the set of all Lebesgue square integrable functions defined on the interval $[0, T],\; \underline{\gamma_{H_i}}   \leq  \gamma_{H_i}(t) \leq  \overline{\gamma_{H_i}}  ,\; 0\leq t \leq T $, where $\overline{\gamma_{H_i}}$  is the upper bound of $\gamma_{H_i}$ and $\underline{\gamma_{H_i}}$  is the lower bound of $\gamma_{H_i}$.

\begin{figure}[!h]
\centering
\includegraphics[width=3.1in]{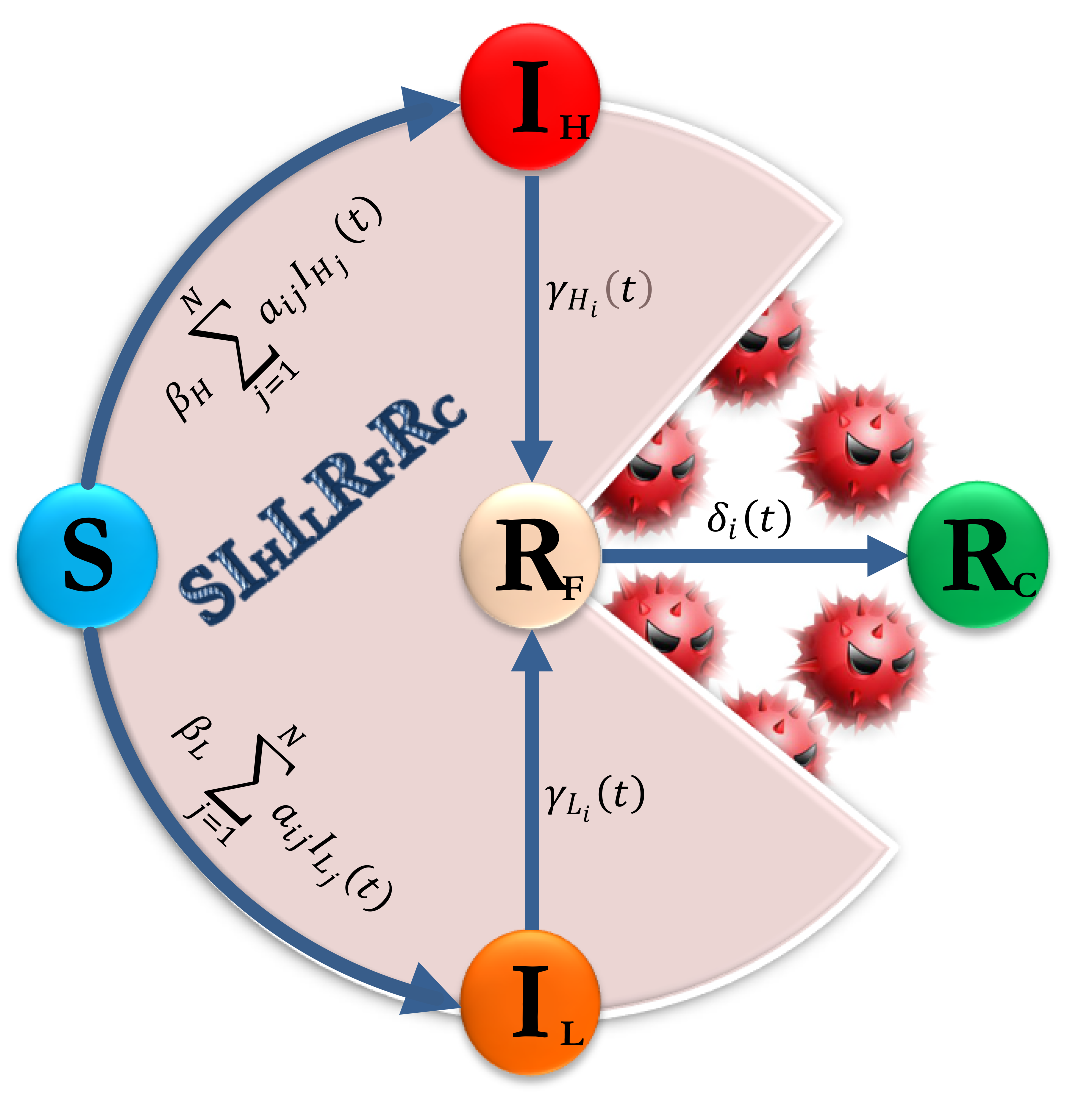}
\caption{ State transition in the $ \mathbf{SI_HI_LR_FR_C}$  model.} 
\label{fig_1a}
\end{figure}

{\textbf{(A4) Infected Low to Recovery First}} $\mathbf{(I_{L_i}(t) \rightarrow R_{F_i}(t))}$:
 at any time, due to first respond to cybersecurity incidents and some sufficient defence mechanisms that used to deal with malware at the early stage in IoT network, the IoT device infected low transit into recovery first stage at an average rate:
\begin{equation}
\label{deqn_ex1a}
\gamma_{L_i}(t),  \qquad 1\leq i \leq N,
\end{equation} 
where $\gamma_{L_i}$ is the rate of applying the security measures that assist the compromised IoT devices to reduce the impact of the cybersecurity attack over IoT network. (Restricted environment by prevent broadcast traffic for the infected devices and other security defence mechanisms).\newline
$\gamma_{L_i}(t) \in L^2[0, T]$, The symbol $L^2[0, T]$ stands for the set of all Lebesgue square integrable functions defined on the interval $[0, T] \; , \underline{\gamma_{L_i}}   \leq  \gamma_{L_i}(t) \leq   \overline{\gamma_{L_i}}  , \; 0\leq t \leq T $, where $\overline{\gamma_{L_i}}$  is the upper bound of $\gamma_{L_i}$ and $\underline{\gamma_{L_i}}$  is the lower bound of $\gamma_{L_i}$.\newline

{\textbf{(A5) Recovery First to Recovery Complete}} $ \mathbf{(R_{F_i}(t)}$ \newline $\rightarrow \mathbf{R_{C_i}(t))}$: 
 at any time, due to the powerful and latest cybersecurity patch, each infected IoT device $i$ will get a latest patch to transit from recovery first into recovery complete stage to become more secure at an average rate:
 \begin{equation}
\label{deqn_ex1a}
\delta_{i}(t),   \qquad 1\leq i \leq N, 
\end{equation}
where $\delta_{i}$ is the recovery rate (powerful and latest security patch). $\delta_{i}(t)\in L^2[0, T]$, The symbol $L^2[0, T]$ stands for the set of all Lebesgue square integrable functions defined on the interval $[0, T],\; \underline{\delta_{i}} \leq  \delta_{i}(t) \leq  \overline{\delta_{i}} ,\; 0\leq t \leq T $, 
where $\overline{\delta_{i}}$  is the upper bound of $\delta_{i}$ and $\underline{\delta_{i}}$  is the lower bound of $\delta_{i}$.

{\autoref{fig_1a}} also describes the probabilistic assumptions of our model, which explain the transition process between IoT devices. 

Here, $Y{\tiny{i}(t)}$ denotes the states of the $ \mathbf{SI_HI_LR_FR_C}$ model at time $t$.

\[ Y{\tiny{i}(t)}=
\left \{
\begin{tabular}{cl} 
0, device $i$ & is Susceptible,\\ 
1, device $i$ & is Infected Low,\\
2, device $i$ & is Infected High,\\
3, device $i$ & is Recover First,\\
4, device $i$ & is Recover Complete.
\end{tabular}
\right \}
\]

Then, let $\Delta t$ be a very small time interval and $O(\Delta t)$ be a higher-order infinitesimal. Assumptions (A1)–(A5) imply that the probabilities of state transition of device $i$ satisfy the following relations.
\begin{equation}
        \left\{
      \begin{split}
             & Pr[Yi(t+\Delta t)=1 \mid Yi(t)=0] \\
             &  \qquad \qquad \qquad  =  \Delta t \; \beta_{L}\; \sum_{j=1}^{N} a_{ij} \;  I_{L_j}(t) +O(\Delta t) ,\\      
             & Pr[Yi(t+\Delta t)=2 \mid Yi(t)=0] \\
             & \qquad \qquad \qquad = \Delta t \; \beta_{H} \; \sum_{j=1}^{N} a_{ij} \; I_{H_j}(t) +O(\Delta t) ,\\
             &Pr[Yi(t+\Delta t)=3 \mid Yi(t)=1]= \Delta t \; \gamma_{L_i}(t)+O(\Delta t) ,\\
             & Pr[Yi(t+\Delta t)=3 \mid Yi(t)=2]= \Delta t \; \gamma_{H_i}(t)+O(\Delta t) , \\ 
             & Pr[Yi(t+\Delta t)=4 \mid Yi(t)=3]= \Delta t  \;\delta_{i}(t)+O(\Delta t). \\ 
       \end{split} 
       \right.
\end{equation} 

The interaction between IoT devices is represented by the following system including of five delayed differential equations. 

Let $\Delta t \rightarrow 0 $ the controlled node-level model can be derived as:

\begin{equation}
      \left\{
       \begin{aligned}
               {\frac{dS_i(t)}{dt}} &= - \beta_{H} \; S_i(t) \;\sum_{j=1}^{N} a_{ij} \;I_{H_j}(t) - \beta_{L} \; S_i(t) \;\sum_{j=1}^{N} a_{ij} \;I_{L_j}(t),\\ 
               &\quad t \geq 0, \; 1\leq i \leq N, \\
               {\frac{dI_{H_i}(t)}{dt}} &= \beta_{H}\; S_i(t) \; \sum_{j=1}^{N} a_{ij} \; I_{H_j}(t) -\gamma_{H_i}(t)I_{H_i}(t) ,\\
               &\quad  t \geq 0, \; 1\leq i \leq N, \\
               {\frac{dI_{L_i}(t)}{dt}} &=  \beta_{L} \; S_i(t)\; \sum_{j=1}^{N} a_{ij} \;  I_{L_j}(t) -\gamma_{L_i}(t)I_{L_i}(t) , \\
               &\quad t \geq 0, \; 1\leq i \leq N, \\
               {\frac{dR_{F_i}(t)}{dt}} & =  \gamma_{H_i}(t)I_{H_i}(t) + \gamma_{L_i}(t)I_{L_i}(t) - \delta_{i}(t)R_{F_i}(t) ,\\
               &\quad t \geq 0, \;  1\leq i \leq N, \\
               {\frac{dR_{C_i}(t)}{dt}} &= \delta_{i}(t)R_{F_i}(t), \qquad \qquad  t \geq 0, \;  1\leq i \leq N. \\
        \end{aligned}
		\right.
    \end{equation}
with the following initial condition:
\begin{equation*}
       \begin{split}
		    & (S_1(0),…,S_N(0),I_{H_1}(0),…,I_{H_N}(0),I_{L_1}(0),…,I_{L_N}(0),\\
                &\quad R_{F_1}(0),…,R_{F_N}(0),R_{C_1}(0),…,R_{C_N}(0))^T \qquad \in \Omega. \\
        \end{split}
    \end{equation*}

where 
\begin{equation*}
      \begin{split}
		  & \Omega = (S_1,…,S_N,I_{H_1},…,I_{H_N},I_{L_1},…,I_{L_N},\\
            &\quad R_{F_1},…,R_{F_N},R_{C_1},…,R_{C_N})^T,  \qquad  i= 1,2,....,N.\\ 
        \end{split}
	\end{equation*}
 
Based on $S_i(t)+I_{H_i}(t)+I_{L_i}(t)+R_{F_i}(t)+R_{C_i}(t)=1 ,$ \; system $(10)$ can by reduced from 5N-dimensional system to 4N-dimensional system. Consequently, the vector $ \mathbf{E}(t) =(  \mathbf{S}(t),  \mathbf{I_{H}}(t), \mathbf{I_{L}}(t), \mathbf{R_{F}}(t))$ 
\begin{equation}
      \begin{split}
		  &= (S_1(t),…,S_N(t),I_{H_1}(t),…,I_{H_N}(t),\\ 
            &\quad I_{L_1}(t),…,I_{L_N}(t), R_{F_1}(t),…,R_{F_N}(t)),\\ 
        \end{split}
	\end{equation}
represents the expected state of the IoT network at time $t$ as following :
\begin{equation}
      \left\{
       \begin{aligned}
           {\frac{dS_i(t)}{dt}} &= - \beta_{H}  S_i(t) \sum_{j=1}^{N} a_{ij} I_{H_j}(t) - \beta_{L}  S_i(t) \sum_{j=1}^{N} a_{ij} I_{L_j}(t),\\
           &\quad  t \geq 0, \; 1\leq i \leq N,\\
            {\frac{dI_{H_i}(t)}{dt}} & = \beta_{H} \; S_i(t) \; \sum_{j=1}^{N} a_{ij} \; 
            I_{H_j}(t) -\gamma_{H_i}(t)I_{H_i}(t),\\
            &\quad t \geq 0, \; 1\leq i \leq N, \\         
           {\frac{dI_{L_i}(t)}{dt}}  & =  \beta_{L} \; S_i(t) \; \sum_{j=1}^{N} a_{ij} \;  I_{L_j}(t) -\gamma_{L_i}(t)I_{L_i}(t), \\
           &\quad t \geq 0, \; 1\leq i \leq N, \\
            {\frac{dR_{F_i}(t)}{dt}}&=  \gamma_{H_i}(t)I_{H_i}(t) + \gamma_{L_i}(t)I_{L_i}(t) - \delta_{i}(t)R_{F_i}(t),\\
            &\quad t \geq 0, \;  1\leq i \leq N, \\
        \bf{\mathbf{E}(0) =\mathbf{E}_0}, \\
       \end{aligned}
		\right.
\end{equation}
initiated with the vector $ \mathbf{E}(0)$ as the initial condition.

 \section{Optimal control strategy}
 \label{Optimal}
This section explains how to apply optimal control theory to analyse the optimal control strategy for mitigating malware propagation. It aims to investigate the existence of an optimal control system that ensures an optimal solution for any given problem. Additionally, the section discusses the determination of the optimality system associated with an optimal control problem.

 The main objective is to find a control $\mathbf{u}(t) \in \mathbf{U}$  where $ {\bf{\mathbf{u}}}= (\delta_1,...,\delta_N, \; \gamma_{H_1},...,\gamma_{H_N}, \; \gamma_{L_1},...,\gamma_{L_N})^T$ so that the cumulative trade-off, denoted $J(\mathbf{u}(t))$, which is defined as the sum of the total number of infected IoT devices in the network during the period time and the cumulative cost of the restricted environment combined with the latest cybersecurity patch. For a basic introduction to the theory of optimal control see \cite{kirk2004optimal,liberzon2011calculus,yang2016optimal}. 

The proposed model aims to investigate the impact of control strategies on several key factors, including: 
\begin{itemize}
\item The number of IoT devices that are infected during the high infection stage.
\item The costs associated with implementing a restricted environment and other security defences during the early recovery stage to combat malware.
\item The costs associated with applying the latest and most robust cybersecurity patch during the complete recovery stage. 
\end{itemize} 

By studying these factors, the model will provide insights into the effectiveness and efficiency of various control strategies in mitigating the spread of malware and securing IoT networks against cyber threats.
For this purpose, we deﬁne $[ 0, T]$ as the time period when an optimal control strategy is imposed on the system and  $  \mathbf{u}(t), (0\leq t \leq T)$  which are given by: 
 $ \mathbf{u}(t) \in L^2[0, T],  \;  \underline{ \mathbf{u}}\leq   \mathbf{u}(t) \leq \;  \overline{ \mathbf{u}}\;, $ 
 where $\overline{\mathbf{u}}$  is the upper bound of $\mathbf{u}$ and $\underline{\mathbf{u}}$  is the lower bound of $\mathbf{u}$, 
which are expressed as $  \underline{\delta_{i}} \leq  \delta_i(t) \leq  \overline{\delta_{i}},  
  \underline{\gamma_{H_i}} \leq  \gamma_{H_i}(t) \leq \overline{\gamma_{H_i}}, 
  \underline{\gamma_{L_i}} \leq  \gamma_{L_i}(t) \leq  \overline{\gamma_{L_i}}, \;   
 1\leq i \leq N, \; \; 0\leq t \leq T. $

 Then, the system represented by equation (12) can be rewritten as follows:
\begin{equation}
         {\frac{d\mathbf{E}(t)}{dt}} = f( \mathbf{E}(t), \mathbf{u}(t)),  \qquad \qquad 0\leq t \leq T.
   \end{equation}
   with the initial condition $\mathbf{E}(0)\in \Omega.$

Subsequently, equation (14) is formulated with the objective of minimising the  total number of infected IoT devices, the costs associated with the restricted environment and other security defences, and the costs associated with the most powerful and up-to-date cybersecurity patch, as follows: \begin{equation*}
         \underset{ \mathbf{u} \in  \mathbf{U}}{\text{Minimize}} \qquad J( \mathbf{u})= \int_{0}^{T} L( \mathbf{E}(t), \mathbf{u}(t))dt. 
   \end{equation*}
   
subject to system (12) or (13) , where
\begin{equation}
         L(\mathbf{E},\mathbf{u})= \sum_{i=1}^{N} \left[I_{H_i} \; + \frac{1}{2} (\delta_{i}^{2})\; + \frac{1}{2} (\gamma_{H_i}^{2} \;+ \gamma_{L_i}^{2})-R_{C_i}\right] 
    \end{equation}

Each $ \mathbf{SI_HI_LR_FR_C}$ model is represented as a 15-tuple consisting of the following elements:

          \begin{equation}
          \begin{split}
               & \mathbb{M}= (G, N,  \beta_{H},\beta_{L} ,\gamma_{H} ,\gamma_{L}, \delta, \overline{\gamma_{L}}, \underline{\gamma_{L}},\overline{\gamma_{H}},\underline{\gamma_{H}} ,\overline{\delta}, \underline{\delta},T ,\bf{E_0}).\\
               \end{split}
              \end{equation}  
\subsection{Existence of an optimal control}
To ascertain the presence of an optimal control for any given problem, it is essential that six specific conditions are met. If the following six conditions are satisfied simultaneously, then there is an optimal control of the problem.\newline 
{\bf{[Condition-1]}} There is \; $  \mathbf{u}(t) \in  \mathbf{U}$   \; such that system (12) is solvable. \newline
{\bf{[Condition-2]}} $ \mathbf{U}$ is convex. \newline
{\bf{[Condition-3]}} $ \mathbf{U}$ is closed.\newline
{\bf{[Condition-4]}} $f( \mathbf{E}, \mathbf{u})$ is bounded by a linear function in $ \mathbf{E}$ .\newline
{\bf{[Condition-5]}} $L( \mathbf{E}, \mathbf{u})$ is convex on $ \mathbf{U}$. \newline
{\bf{[Condition-6]}} $L( \mathbf{E}, \mathbf{u}) \geq  c_1 \parallel {\bf{ \mathbf{u}}} \parallel_2 ^{\rho}  + c_2 \; for \; some \; vector\; norm \newline \parallel \bullet \parallel, {\rho} >1 , \; c_1> 0 \; and \; c_2 . $

{\autoref{Appendixa} provides more details on the fulfilment of the six conditions.

\subsection{The optimality system}
The optimality system associated with an optimal control problem is recognised as providing a numerical solution. To achieve this goal, we focus on the corresponding Hamiltonian function.

\begin{equation*}
	H(\mathbf{E}(t),\mathbf{u}(t),\lambda(t))= L(\mathbf{E},\mathbf{u}) + \lambda(t)\; f(\mathbf{E},\mathbf{u}). 
\end{equation*}
\begin{equation*}
      \begin{split}
		  & H(\mathbf{E}(t),\mathbf{u}(t),\lambda(t))= L(\mathbf{E},\mathbf{u})  \\
            & \qquad \qquad + \sum_{i=1}^{N} \lambda^{S}_{i} \left[{\frac{dS_i(t)}{dt}}\right]  + \sum_{i=1}^{N} \lambda^{H}_{i} \left[{\frac{dI_{H_i}(t)}{dt}}\right] \\
            &\qquad  \qquad  + \sum_{i=1}^{N} \lambda^{L}_{i} \left[{\frac{dI_{L_i}(t)}{dt}}\right]+ \sum_{i=1}^{N} \lambda^{F}_{i} \left[{\frac{dR_{F_i}(t)}{dt}}\right]. \\ 
        \end{split}
	\end{equation*}
The final expression of the Hamiltonian function takes the form of the following equation
\begin{equation}
      \begin{split}
            & H(\mathbf{E}(t),\mathbf{u}(t),\lambda(t))=\\
            & \quad \sum_{i=1}^{N} \Biggr[I_{H_i} + \frac{1}{2} \delta_{i}^{2} + \frac{1}{2} (\gamma_{H_i}^{2} + \gamma_{L_i}^{2}) -R_{C_i}\Biggr] \\ 
            & + \sum_{i=1}^{N} \lambda^{S}_{i} \Biggr[ - \beta_{H} \; S_i(t)  \sum_{j=1}^{N} a_{ij} I_{H_j}(t) - \beta_{L} \; S_i(t) \;\sum_{j=1}^{N} a_{ij} \;  I_{L_j}(t)\Biggr]\\
            & + \sum_{i=1}^{N} \lambda^{H}_{i} \Biggr[ \beta_{H} \; S_i(t) \; \sum_{j=1}^{N} a_{ij} \; I_{H_j}(t) -\gamma_{H_i}(t)I_{H_i}(t)\Biggr]\\
            & + \sum_{i=1}^{N} \lambda^{L}_{i} \Biggr[ \beta_{L} \; S_i(t) \; \sum_{j=1}^{N} a_{ij} \;  I_{L_j}(t) -\gamma_{L_i}(t)I_{L_i}(t)\Biggr]\\
             & + \sum_{i=1}^{N} \lambda^{F}_{i} \Biggr[ \gamma_{H_i}(t)I_{H_i}(t) + \gamma_{L_i}(t)I_{L_i}(t) - \delta_{i}(t)R_{F_i}(t)\Biggr].\\
             &0\leq t \leq T ,\; \; 1\leq i \leq N,\\
        \end{split}
	\end{equation}
where  
$\lambda= (\lambda^{S}_{1},...,\lambda^{S}_{N}, \lambda^{H}_{1},...,\lambda^{H}_{N}, \lambda^{L}_{1},...,\lambda^{L}_{N}, \lambda^{F}_{1},...,\lambda^{F}_{N})^T$ is the adjoint.\newline

Assume that $\mathbf{u}$ represents the optimal control for the $ \mathbf{SI_HI_LR_FR_C}$ model (15), and $\mathbf{E}$ denotes the solution of the associated controlled model (14). Subsequently, there exist an adjoint function ($\lambda$) such the following equation.

 \begin{equation*}
       \left\{
      \begin{split}
             & \frac{d \lambda^{S}_{i}}{dt} = -\frac{ \partial H} {\partial S_i} = \lambda^{S}_{i} \; \beta_{H} \;\sum_{j=1}^{N} a_{ij} \;I_{H_j} + \lambda^{S}_{i} \; \beta_{L} \;\sum_{j=1}^{N} a_{ij} \;  I_{L_j} \\
             & \qquad \qquad \qquad - \lambda^{H}_{i} \; \beta_{H} \; \sum_{j=1}^{N} a_{ij} \; I_{H_j}   - \lambda^{L}_{i} \; \beta_{L} \; \sum_{j=1}^{N} a_{ij} \;  I_{L_j} \; , \\
             &=\Bigr[\beta_{H}\sum_{j=1}^{N} a_{ij} I_{H_j}\Bigr]\Bigr[\lambda^{S}_{i} -\lambda^{H}_{i}\Bigr]+ \Bigr[ \beta_{L}\sum_{j=1}^{N}a_{ij} I_{L_j} \Bigr] \Bigr[\lambda^{S}_{i}-\lambda^{L}_{j}\Bigr]  \\
             & \frac{d \lambda^{H }_{i}}{dt}= -\frac{ \partial H}  {\partial I_{H_i}} = -1 + \beta_{H}\;\sum_{j=1}^{N} a_{ij}  \;S_j  \;\lambda^{S}_{j} + \lambda^{H}_{i} \; \gamma_{H_i} \\
             & \qquad \qquad \qquad  - \beta_{H}\; \sum_{j=1}^{N} a_{ij}  \;S_j  \;\lambda^{H}_{j} - \;\lambda^{F}_{i}\; \gamma_{H_i} \; , \\
             &  =  -1 + \beta_{H} \; \sum_{j=1}^{N} a_{ij} \;  S_j \;  \Bigr[  \lambda^{S}_{j} - \lambda^{H}_{j} \Bigr] +  \gamma_{H_i} \Bigr[\lambda^{H}_{i}- \lambda^{F}_{i}  \Bigr]  \; ,  \\
             &  \frac{d \lambda^{L}_{i}}{dt}= -\frac{ \partial H}  {\partial I_{L_i}} =  \;\beta_{L}  \;\sum_{j=1}^{N} a_{ij} \; S_j \;\lambda^{S}_{j} \;  +  \lambda^{L}_{i} \; \gamma_{L_i}\\ 
             & \qquad \qquad \qquad - \;\beta_{L}\; \sum_{j=1}^{N} a_{ij} \; S_j \;\lambda^{L}_{j} -\lambda^{F}_{i} \;\gamma_{L_i} \; , \\
             &  =   \beta_{L} \;\sum_{j=1}^{N} a_{ij} \; S_j \; \Bigr[  \lambda^{S}_{j} - \lambda^{L}_{j} \Bigr]  + \gamma_{L_i}\Bigr[ \lambda^{L}_{i} - \lambda^{F}_{i} \Bigr]  \; ,  \\
             & \frac{d \lambda^{F }_{i}}{dt}= - \frac{ \partial H}  {\partial R_{F_i}} = \lambda^{F}_{i} \;\delta_{i} \; . \\
       \end{split} 
       \right.
    \end{equation*} 
    According to the Pontryagin Minimum Principle  \cite{fleming2012deterministic,kamien2012dynamic,fister1998optimizing,nowzari2015optimal} there exists functions $\; \lambda^{S \ast}_{i} (t), \lambda^{H \ast}_{i} (t), \lambda^{L \ast}_{i} (t), \lambda^{F \ast}_{i} (t) \; \; \; 0\leq t \leq T ,\; \; 1\leq i \leq N,$ such that 
\begin{equation}
      \begin{split}
             &  \frac{d \lambda^{S \ast}_{i}}{dt} = - \frac{ \partial H( \mathbf{E}^{\ast}(t), \lambda^{\ast}(t),  \mathbf{u}^{\ast}(t))}  {\partial S_i(t)}, \\
             &  \frac{d \lambda^{H \ast}_{i}}{dt} = -  \frac{ \partial H( \mathbf{E}^{\ast}(t), \lambda^{\ast}(t),  \mathbf{u}^{\ast}(t))}  {\partial I_{H_i}(t)}, \\
             &  \frac{d \lambda^{L \ast}_{i}}{dt} = - \frac{ \partial H( \mathbf{E}^{\ast}(t), \lambda^{\ast}(t),  \mathbf{u}^{\ast}(t))}  {\partial I_{L_i}(t)}, \\
             &  \frac{d \lambda^{F \ast}_{i}}{dt} = -  \frac{ \partial H( \mathbf{E}^{\ast}(t), \lambda^{\ast}(t),  \mathbf{u}^{\ast}(t))}  {\partial R_{F_i}(t)}, \\
       \end{split} 
   \end{equation}  
 Then we can obtain the following optimal system of equations:
\begin{equation}
       \left\{
       \begin{split}
             & {\frac{dS_i(t)}{dt}} = - \beta_{H} \; S_i(t) \; \sum_{j=1}^{N} a_{ij} \;I_{H_j}(t)  - \beta_{L} \; S_i(t) \;\sum_{j=1}^{N} a_{ij} \;  I_{L_j}(t),\\
             & \qquad \qquad t \geq 0, \; 1\leq i \leq N, \\
             & {\frac{dI_{H_i}(t)}{dt}} = \beta_{H} \; S_i(t) \; \sum_{j=1}^{N} a_{ij} \; I_{H_j}(t) -\gamma_{H_i}(t)I_{H_i}(t),\\
             & \qquad \qquad t \geq 0, \; 1\leq i \leq N, \\
             & {\frac{dI_{L_i}(t)}{dt}} =  \beta_{L} \; S_i(t) \; \sum_{j=1}^{N} a_{ij} \;  I_{L_j}(t) -\gamma_{L_i}(t)I_{L_i}(t),\\
             & \qquad \qquad t \geq 0, \; 1\leq i \leq N, \\
             & {\frac{dR_{F_i}(t)}{dt}} =  \gamma_{H_i}(t)I_{H_i}(t) + \gamma_{L_i}(t)I_{L_i}(t) - \delta_{i}(t)R_{F_i}(t),\\
             & \qquad \qquad t \geq 0, \; 1\leq i \leq N, \\
             & \frac{d \lambda^{S}_{i}}{dt} =   \Bigr[ \beta_{H}\;\sum_{j=1}^{N} a_{ij} \;I_{H_j} \Bigr]\Bigr[\lambda^{S}_{i} -\lambda^{H}_{i}\Bigr]  +   \Bigr[ \beta_{L}\;\sum_{j=1}^{N} a_{ij} \;  I_{L_j}   \Bigr] \\
             &\qquad \qquad \Bigr[\lambda^{S}_{i}- \lambda^{L}_{j}\Bigr],  \; t \geq 0, \; 1\leq i \leq N,\\
             &  \frac{d \lambda^{H }_{i}}{dt}= -1 +  \beta_{H} \; \sum_{j=1}^{N} a_{ij} \;  S_j \;  \Bigr[  \lambda^{S}_{j} - \lambda^{H}_{j} \Bigr]\\
             &\qquad \qquad +  \gamma_{H_i} \Bigr[\lambda^{H}_{i}- \lambda^{F}_{i}  \Bigr], \qquad t \geq 0, \; 1\leq i \leq N,\\
             &  \frac{d \lambda^{L}_{i}}{dt}=   \beta_{L} \;\sum_{j=1}^{N} a_{ij} \; S_j \; \Bigr[  \lambda^{S}_{j} - \lambda^{L}_{j} \Bigr]  + \gamma_{L_i}\Bigr[ \lambda^{L}_{i} - \lambda^{F}_{i} \Bigr] , \\
             & \qquad \qquad t \geq 0, \; 1\leq i \leq N, \\
             &   \frac{d \lambda^{F }_{i}}{dt} = \lambda^{F}_{i} \;\delta_{i},\qquad  t \geq 0, \; 1\leq i \leq N. \\
             &\delta_{i}(t)= \max \{ \min \{ \lambda^{F}_{i}(t) *R_{F_i}(t), \; \overline{\delta_{i}} \} ,\;\underline{\delta_i} \}, \\
             & \qquad \qquad t \geq 0, \; 1\leq i \leq N, \\
             &\gamma_{H_i}(t)=\max \{ \min \{(\lambda^{H}_{i}(t)- \lambda^{F}_{i}(t)) *I_{H_i}(t),\; \overline{\gamma_{H_i}} \} ,\underline{\gamma_{H_i}} \}, \\
             & \qquad \qquad t \geq 0, \; 1\leq i \leq N, \\
             &\gamma_{L_i}(t)=\max \{ \min \{(\lambda^{L}_{i}(t)- \lambda^{F}_{i}(t)) *I_{L_i}(t), \; \overline{\gamma_{L_i}} \} ,\underline{\gamma_{L_i}} \}, \\
             & \qquad \qquad t \geq 0, \; 1\leq i \leq N, \\
             & \bf{ \mathbf{E}(0) = \mathbf{E}_0}.\\
        \end{split} 
        \right.
   \end{equation}

\begin{figure*}[!t]
        \centering
         \includegraphics[width=6.8in,height=3.5in]{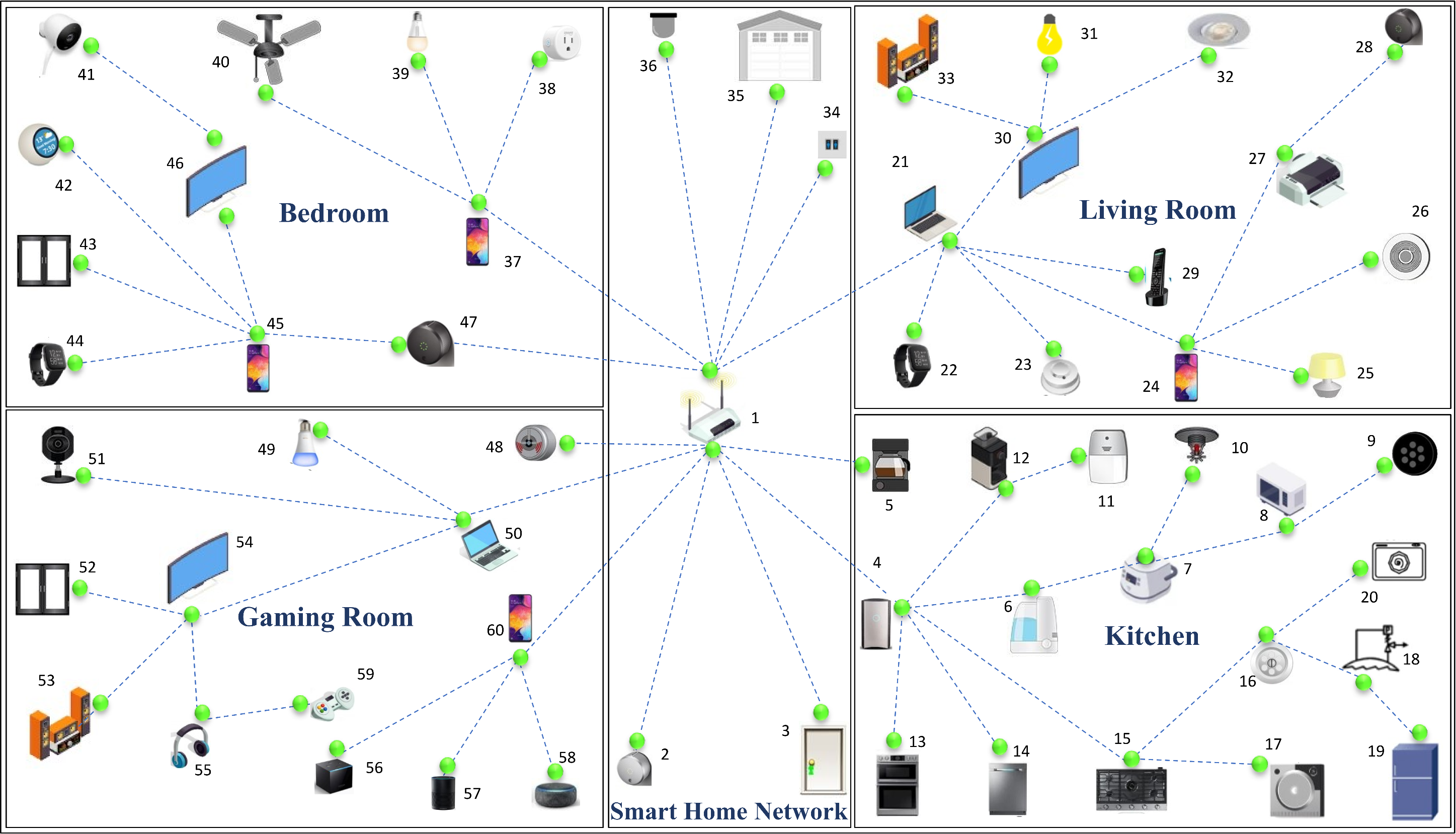}
         \vspace*{-2mm}
         \caption{The smart home dataset.} 
         \label{fig_2a}
    \end{figure*}

\section{Experiments and analysis of the results}
\label{Experiments}
The purpose of this section is to elucidate the experiments that will serve as the focal point of our paper through the use of numerical simulations. Our experimentation will follow three objectives. Initially, the effectiveness of the optimal control strategy will be verified. Secondly, the $ \mathbf{SI_HI_LR_FR_C}$ model will be examined in two scenarios: one with immediate response state and the other without immediate response state, within the context of the restricted environment, as part of the optimal control strategy analysis. Finally, the impact of the malware on the IoT devices through a high rate of infection and a low rate of infection will be examined under the umbrella of optimal control strategy.

In the scope of the current experiments, we have employed the advanced computational capabilities of Matlab R2022a to simulate the optimal control strategy for the $ \mathbf{SI_HI_LR_FR_C}$ model. The optimal control strategy was obtained by employing the highly efficient forward-backward sweep method, a widely utilized approach \cite{kandhway2016optimal, asano2008optimal, gaff2009optimal, kandhway2014run}. 
Then, the $ \mathbf{SI_HI_LR_FR_C}$ model is utilized for all experiments.
          \begin{equation*}
          \begin{split}
               & \mathbb{M}= (G, N,  \beta_{H},\beta_{L} ,\gamma_{H} ,\gamma_{L}, \delta, \overline{\gamma_{L}}, \underline{\gamma_{L}},\overline{\gamma_{H}},\underline{\gamma_{H}} ,\overline{\delta}, \underline{\delta},T ,\bf{E_0}).\\
               \end{split}
              \end{equation*}

\subsection{The smart home dataset} 
For our purpose, the dataset that was used in the experiments was created to be a simulation of the real-world scenario for IoT networks in the smart home. This dataset consists of sixty IoT devices that represent the IoT network for the smart home, which is a sophisticated system consisting of five key components, namely the living room, kitchen, gaming room, and bedroom. Each component is seamlessly integrated into the network to enable smart and automated functionalities that enhance the convenience and comfort of modern living. The living room serves as a central hub for entertainment and home automation control, while the kitchen incorporates IoT devices for efficient cooking, food management, and energy monitoring. The gaming room is equipped with smart gaming consoles and accessories, and the bedroom is equipped with IoT devices for enhanced comfort, security, and sleep tracking. Together, these components form a comprehensive IoT network that transforms a regular home into a cutting-edge smart home as shown in  {\autoref{fig_2a}}.

\begin{algorithm}
     \caption{\textbf{$ \mathbf{SI_HI_LR_FR_C}$ Optimization}}
     \hspace*{0.02in} {\bf Input:} an instance of the $\mathbf{SI_HI_LR_FR_C}$  model (18),  
     \hspace*{0.26in} denoted by $\mathbb{M}=(G, N,  \beta_{H},\beta_{L} ,\gamma_{H} ,\gamma_{L}, \delta, \overline{\gamma_{L}}, \underline{\gamma_{L}},$
     \hspace*{0.26in}$\overline{\gamma_{H}},\underline{\gamma_{H}} ,\overline{\delta}, \underline{\delta},T ,\bf{ \mathbf{E}_0})$, $n$ the number of iterations, convergence error $\epsilon .$ \newline 	
     \hspace*{0.02in} {\bf Output:} The best optimal control strategy $(\mathbf{u})$ and the state evolution $(\mathbf{x})$ associated with $(\mathbf{u})$. 
     \begin{algorithmic}[1] 
            \State $ k \leftarrow 0$;  $\mathbf{x}^{(0)} \leftarrow 0 ;$ $\mathbf{u}^{(0)} \leftarrow 0 ;$
            \Repeat 
		\State $k  \leftarrow k+1  $;
            \State  Based on system (12) with  $\mathbf{x} \leftarrow \mathbf{x}^{(k-1)}, \mathbf{u}\leftarrow \mathbf{u}^{(k-1)},$ \hspace*{0.16 in} and $ \bf{E}(0)\leftarrow \bf{E_0}$, forwardly calculate the states \hspace*{0.16 in} function $\bf{E}$;
            \State  $ \bf{E}^{(k)} \leftarrow \bf{E};$
  	  \State  Based on system (18) with $\mathbf{x}\leftarrow x^{(k-1)}, \mathbf{u}\leftarrow \mathbf{u}^{(k-1)},$ \hspace*{0.18 in}$ \bf{E}\leftarrow \bf{E^{(k)}}$, and $ \lambda^{S}(T)= \lambda^{L_H}(T)= \lambda^{I_L}(T)=\hspace*{0.16 in} \lambda^{R_F}(T)=0$, backwardly calculate the adjoint \hspace*{0.16 in} function $ \lambda^{S}, \lambda^{L_H}, \lambda^{I_L},\lambda^{R_F}$;
            \State  $\lambda^{S(k)} \leftarrow \lambda^{S}; \lambda^{I_H(k)}\leftarrow \lambda^{I_H}; \lambda^{I_L(k)} \leftarrow \lambda^{I_L}; \lambda^{R_F(k)} \leftarrow \lambda^{R_F};$
            \State Based on system (18) with $ \bf{E}\leftarrow\bf{E^{(k)}}$,  $\lambda^{S}\leftarrow \lambda^{S(k)} , \lambda^{I_H}\leftarrow \lambda^{I_H(k)} ,  \lambda^{I_L}\leftarrow\lambda^{I_L(k)} , \lambda^{R_F}\leftarrow\lambda^{R_F(k)} $, calculate $\mathbf{x}$ and $\mathbf{u}$;
            \State $\mathbf{x}^{(k)}\leftarrow \mathbf{x}; \mathbf{u}^{(k)}\leftarrow \mathbf{u} ;$
            \Until {$\parallel \mathbf{x}^{(k)} - \mathbf{x}^{(k-1)}\parallel + \parallel \mathbf{u}^{(k)} - \mathbf{u}^{(k-1)}\parallel < \epsilon$ or $k \geq n$};
 	   \State return $ (\mathbf{x}^{(k)} ,\mathbf{u}^{(k)} ) $;
        \end{algorithmic}
        \label{Alga}
        
\end{algorithm}

\subsection{The proposed algorithm for solving the optimality system:}

The primary objectives of the optimal control strategy are as follows: first, to minimize the cost associated with implementing of the latest and most robust cybersecurity patch during the complete recovery stage. Secondly, to reduce the cost of deploying the restricted environment and other security defense mechanisms during the recovery first stage at an early stage. By achieving these objectives, the optimal control strategy aims to enhance the ability of the IoT network to combat malware effectively which will be an asset in the fight against malware in the IoT network.
The optimization of the $ \mathbf{SI_HI_LR_FR_C}$ model can be achieved by following the sequence code outlined in  Algorithm {\autoref{Alga}}.

To investigate the effectiveness of optimal control strategies in addressing systems of optimality, a controlled experiment will be conducted.\newline \newline

\textbf{Experiment 1:} A system of optimality can be effectively addressed by leveraging an optimal control strategy.

In the present experiment, we have simulated the optimal control strategy for the $ \mathbf{SI_HI_LR_FR_C}$ model to evaluate the effectiveness of the proposed strategy, we conducted a comparison of infected devices in scenarios with and without optimal control. This comparative analysis provides compelling evidence of the considerable impact of the optimal control strategy on the reduction of infected devices. Overall, the findings of this experiment underscore the practical significance of the optimal control strategy and its potential to improve the performance of the $ \mathbf{SI_HI_LR_FR_C}$  model.
 Four different cases have been under consideration.

 {\textbf{Case 1:} Assume that the parameters of the system (18) are as follows: $N=60, \; \beta_{H}=0.0004, \; \beta_{L}=0.0002, \;\gamma_{H_i}=0.6, \; \gamma_{L_i}=0.4, \; \delta_{i}=0.9,\; \overline{\gamma_{L_i}}=0.6, \; \underline{\gamma_{L_i}}=0.1, \;  \overline{\gamma_{H_i}}=1, \underline{\gamma_{H_i}}=0.1 , \; \overline{\delta_{i}}=0.8 , \; \underline{\delta_{i}}=0.1$.\newline

Through solving the optimality system for case 1, the results are shown in  {\autoref{fig_3a}}, which plots four different samples from the optimality system for reducing the number of infected IoT devices, the cost of applying the restricted environment, and the cost of applying the latest cybersecurity patch.

{\textbf{Case 2:} Assume that the parameters of the system (18) are as follows: $N=60, \; \beta_{H}=0.0004, \; \beta_{L}=0.0002, \;\gamma_{H_i}=0.6, \; \gamma_{L_i}=0.4, \; \delta_{i}=0.9,\; \overline{\gamma_{L_i}}=0.3, \; \underline{\gamma_{L_i}}=0.1, \;  \overline{\gamma_{H_i}}=0.5, \underline{\gamma_{H_i}}=0.1 , \; \overline{\delta_{i}}=0.7 , \; \underline{\delta_{i}}=0.1$.

 Through solving the optimality system for case 2, the results are shown in  {\autoref{fig_4a}}, which plots four different samples from the optimality system for reducing the number of infected IoT devices, the cost of applying the restricted environment, and the cost of applying the latest cybersecurity patch.
\begin{figure}[!h]
        \centering
        \includegraphics[width=3.4in,height=3in]{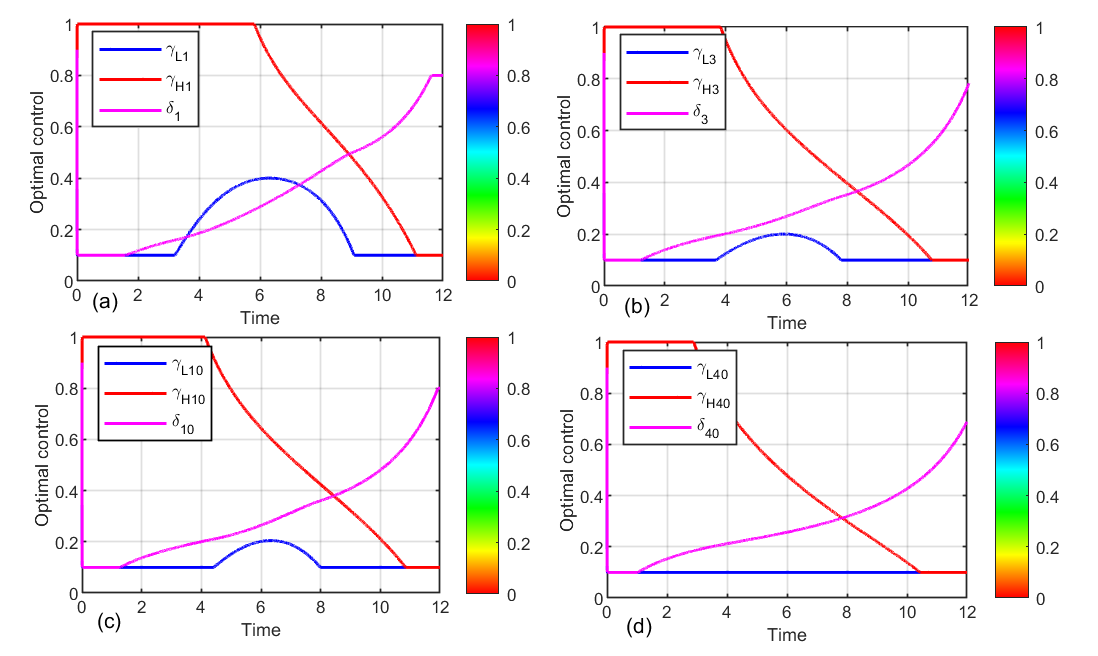}
        \vspace*{-6mm}
        \caption{Samples from the optimality system for case 1.} 
        \label{fig_3a}
\end{figure}

 \begin{figure}[!h]
        \centering
        \includegraphics[width=3.4in,height=3in]{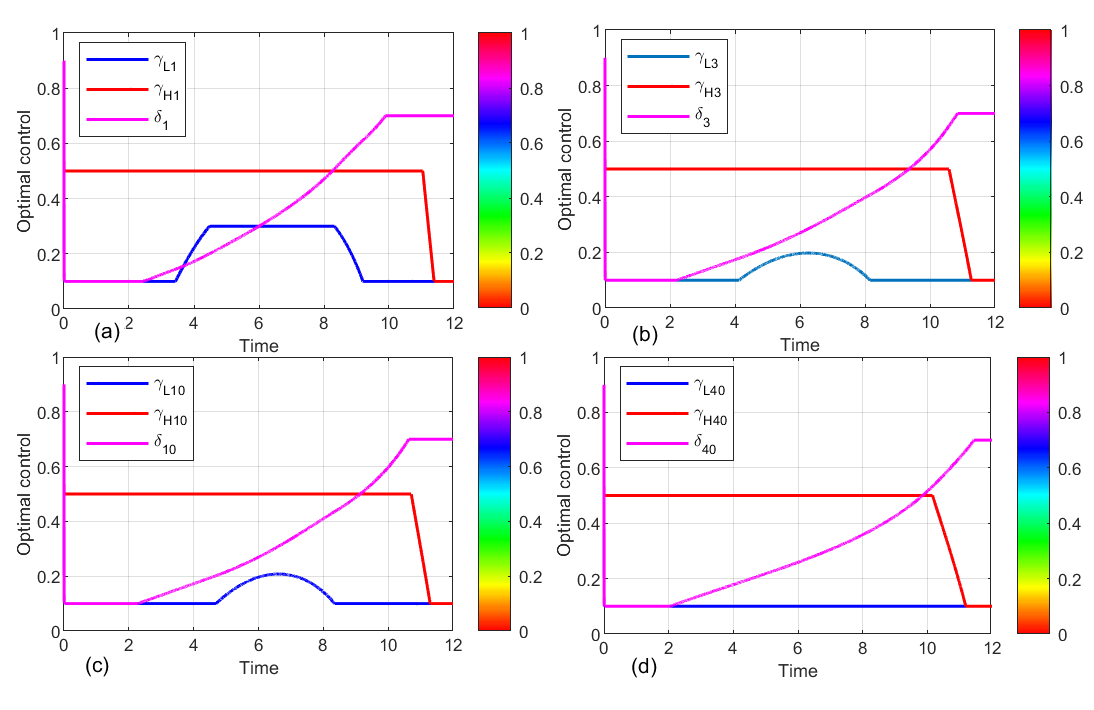}
        \vspace*{-6mm}
        \caption{Samples from the optimality system for case 2.} 
        \label{fig_4a}
\end{figure}

 {\textbf{Case 3:} Assume that the parameters of the system (18) are as follows: $N=60, \; \beta_{H}=0.0006, \; \beta_{L}=0.0004, \;\gamma_{H_i}=0.5, \; \gamma_{L_i}=0.1, \; \delta_{i}=0.9,\; \overline{\gamma_{L_i}}=0.6, \; \underline{\gamma_{L_i}}=0.1, \;  \overline{\gamma_{H_i}}=1, \underline{\gamma_{H_i}}=0.1 , \; \overline{\delta_{i}}=0.8 , \; \underline{\delta_{i}}=0.1$. 

Through solving the optimality system for case 3, the results are shown in  {\autoref{fig_5a}}, which plots four different samples from the optimality system for reducing the number of infected IoT devices, the cost of applying the restricted environment, and the cost of applying the latest cybersecurity patch.

{\textbf{Case 4:} Assume that the parameters of the system (18) are as follows: $N=60, \; \beta_{H}=0.0006, \; \beta_{L}=0.0004, \;\gamma_{H_i}=0.5, \; \gamma_{L_i}=0.1, \; \delta_{i}=0.9,\; \overline{\gamma_{L_i}}=0.3, \; \underline{\gamma_{L_i}}=0.1, \;  \overline{\gamma_{H_i}}=0.5, \underline{\gamma_{H_i}}=0.1 , \; \overline{\delta_{i}}=0.7 , \; \underline{\delta_{i}}=0.1$. 

Through solving the optimality system for case 4, the results are shown in  {\autoref{fig_6a}}, which plots four different samples from the optimality system for reducing the number of infected IoT devices, the cost of applying the restricted environment, and the cost of applying the latest cybersecurity patch.
Furthermore, a series of 100 analogous experiments have been conducted, and uniform observations of the phenomena have been noted across all the experiments.

\begin{figure}[!h]
        \centering
        \includegraphics[width=3.4in,height=3in]{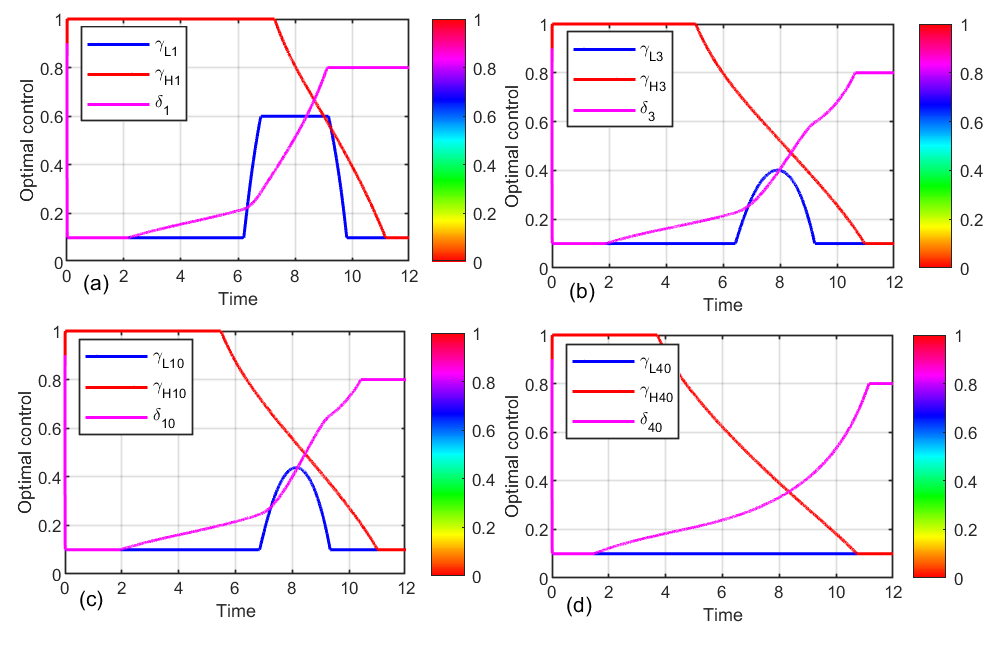}
        \vspace*{-6mm}
        \caption{Samples from the optimality system for case 3.} 
        \label{fig_5a}
    \end{figure}
    
   \begin{figure}[!h]
        \centering
        \includegraphics[width=3.4in,height=3in]{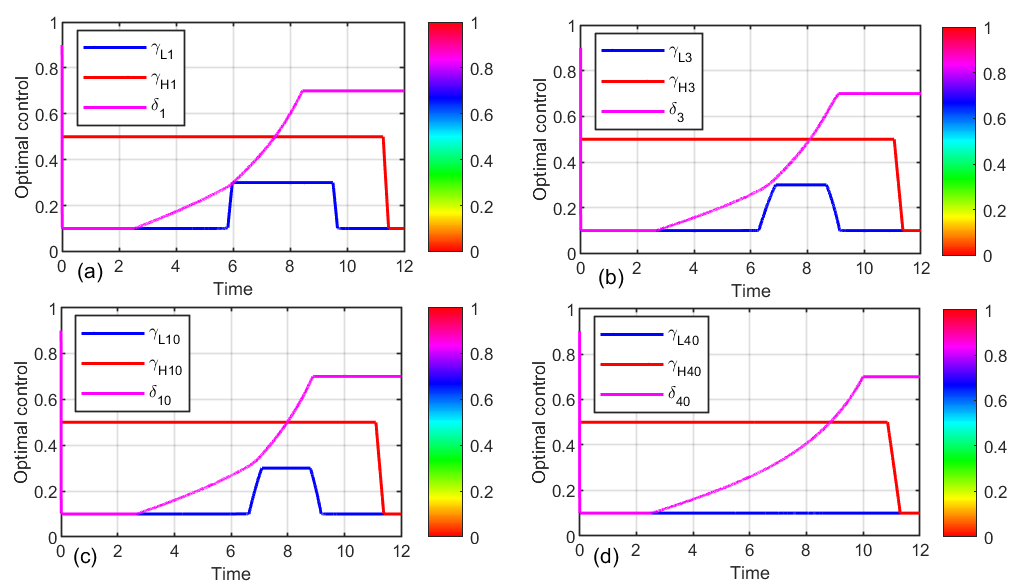}
        \vspace*{-6mm}
        \caption{Samples from the optimality system for case 4.} 
        \label{fig_6a}
    \end{figure}

\begin{table*} [!h]
\centering
\caption{Restricted environment strategy.}
\centering
\begin{tabular}{>{\centering\arraybackslash} m{1cm}>{\centering\arraybackslash} m{4.7 cm}>{\centering\arraybackslash} m{4.7 cm} >{\centering\arraybackslash}m{4.7 cm} }
\hline
 & A non-compromised device & Device with Infected Low ($\mathbf{I_L}$) &Device with Infected High ($\mathbf{I_H}$) \\\hline
Unicast & \checkmark   & \checkmark & \checkmark  \\ 
Multicast & \checkmark & \checkmark & - \\ 
Broadcast & \checkmark & - &  - \\ \hline
\end{tabular}
\label{tab:Table3}
\end{table*}

\newpage

\subsection{The proposed algorithm for randomly generating control strategies:}    
This subsection concerns the generation of control strategies through a randomized approach.
Algorithm \autoref{algb}
 outlines the RGCS algorithm, which stands for randomly generating a control strategy, designed for the purpose of generating a control strategy through random means. The approach involves the random partitioning of the time horizon $[0, T]$ into subintervals, wherein consistent rates are randomly generated for each node and subinterval. In the course of our experiments, we have set the value of $n$ to 100. Our comparative experiment is ready to be conducted.

\begin{algorithm}
     \caption{\textbf{RGCS}}
     \hspace*{0.02in} {\bf Input:} an instance of the $\mathbf{SI_HI_LR_FR_C}$  model (12), denoted by: \newline $\mathbb{M}=(G, N,  \beta_{H},\beta_{L} ,\gamma_{H} ,\gamma_{L}, \delta, \overline{\gamma_{L}}, \underline{\gamma_{L}},$
     $\overline{\gamma_{H}},\underline{\gamma_{H}} ,\overline{\delta}, \underline{\delta},T ,\bf{ \mathbf{E}_0})$,
     positive integer $n$. \newline
     \hspace*{0.02in} {\bf Output:} a control strategy $(\mathbf{u})$.
     \begin{algorithmic}[1] 
            \State randomly and uniformly generate a partition of $[0,T] : 0 =t_0 < t_1< ....<t_n<t_{n+1}=T ; $ 
            \For{$1\leq i \leq N $}
                \For {$0 \leq k \leq n$}
                \State randomly and uniformly generate $\eta \in [0,\overline{S_i}] ,\mu\in \hspace*{0.37 in} [0,\overline{I_{H_i}}], \psi \in [0,\overline{I_{L_i}}], \phi \in [0,\overline{R_{F_i}}], \xi \in [0,\overline{R_{C_i}}];$
                \For{$t_k \leq t \leq t_{k+1}$}
                   \State $S_i(t)\leftarrow \eta;$ $I_{H_i}(t)\leftarrow \mu;$ $ I_{L_i}(t)\leftarrow \psi;  $ \hspace*{0.55in}  $R_{F_i}(t) \leftarrow \phi; $  $R_{C_i}(t) \leftarrow \xi;$
                 \EndFor
  		    \EndFor
                \State $S_i(T)\leftarrow S_i(t_n);$ $I_{H_i}(T)\leftarrow I_{H_i}(t_n);$  $ I_{L_i}(T)\leftarrow \hspace*{0.15in} I_{L_i}(t_n);  $  $R_{F_i}(T) \leftarrow R_{F_i}(t_n); $  $R_{C_i}(T) \leftarrow R_{C_i}(t_n);$         \EndFor
                \State Based on system (12) calculate $(\mathbf{u})$;
 	    \State return $ ( \mathbf{u}) $;
              \end{algorithmic}
        \label{algb}
\end{algorithm}

 {\textbf{Experiment 2:} A comparison is conducted between the optimal control strategy for the $ \mathbf{SI_HI_LR_FR_C}$ model and the RGCS algorithm.
 
In light of executing the RGCS algorithm 100 times, a series of 100 $J$ control strategies is generated. That qualiﬁed $ \mathbf{SI_HI_LR_FR_C}$  policies, denoted $\Upsilon =   \lbrace \mathbf{u}^1, ......, \mathbf{u}^{100} \rbrace $.  As depicted in  {\autoref{fig_7a}}.
$ J(\mathbf{u}), \mathbf{u} \in  \lbrace \mathbf{u}^*  \rbrace  \cup \Upsilon $. It is seen that $J(\mathbf{u}^*) < J(\mathbf{u}), \mathbf{u} \in \Upsilon $. Consequently, the policy $u^*$ is superior to all eligible  $ \mathbf{SI_HI_LR_FR_C}$  policies.
Based on the experiment conducted above and 100 other similar experiments, it has been consistently observed that the $ \mathbf{SI_HI_LR_FR_C}$ control exhibits superior performance.

 \begin{figure}[!h]
\centering
\includegraphics[width=3.5in, height=2.9in]{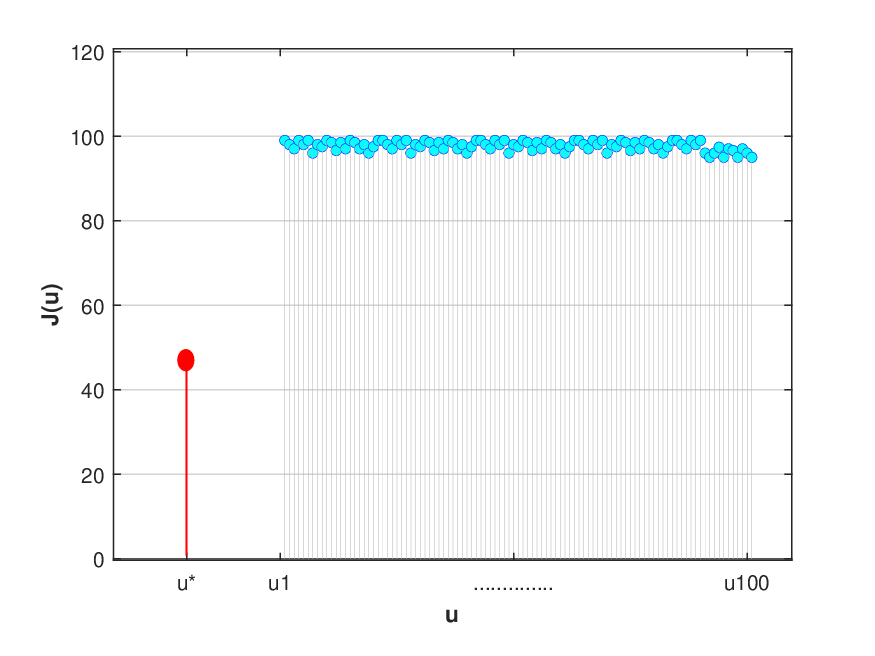}
\vspace*{-6mm}
\caption{ $ J(\mathbf{u}), \mathbf{u} \in  \lbrace \mathbf{u}^*  \rbrace  \cup \Upsilon ,$ in experiment 2.} 
\label{fig_7a}
\end{figure} 

{\textbf{Experiment 3:} A comparison is conducted between the  $ \mathbf{SI_HI_LR_FR_C}$ model with immediate response state and without immediate response state in a restricted environment.}

The strategy of the restricted environment is focused on the effective management of bandwidth traffic types, which include unicast, multicast, and broadcast. By implementing the immediate response state, full control over these traffic types can be maintained, which in turn helps to reduce the propagation of malware within the IoT network. When functioning normally, a device that is not compromised typically generates three types of network traffic that use the available bandwidth: unicast, multicast, and broadcast. However, when a device is infected with high-propagation-rate malware, it generates only unicast traffic, while a device is infected with low-propagation-rate malware causes the device to generate both unicast and multicast traffic, as detailed in \autoref{tab:Table3}.
The preparation and deployment of a new patch can be a time-consuming process due to the diverse types of malware. Moreover, the rapid propagation of malware across the network can exacerbate the situation. By employing a proactive approach through the implementation of the immediate response strategy, guided by the optimal control strategy, it is possible to effectively minimize both inter-device communication and the potential dissemination of malware within IoT networks.

The parameters for applying the $ \mathbf{SI_HI_LR_FR_C}$ model without restricted environment are set as the follows: \newline  
$N=60, \; T=12 \; \beta_{H}=0.004, \; \beta_{L}=0.002, \; \delta_{i}=0.5,\; $.
                  
The depicted data in  {\autoref{fig_8a}} indicates that a high rate of spread has infected around 46 IoT devices, whereas a low rate of spread has infected only 6 IoT devices.   
  \begin{figure}[!h]
         \centering
          \includegraphics[width=3.2in  , height=3.0in]{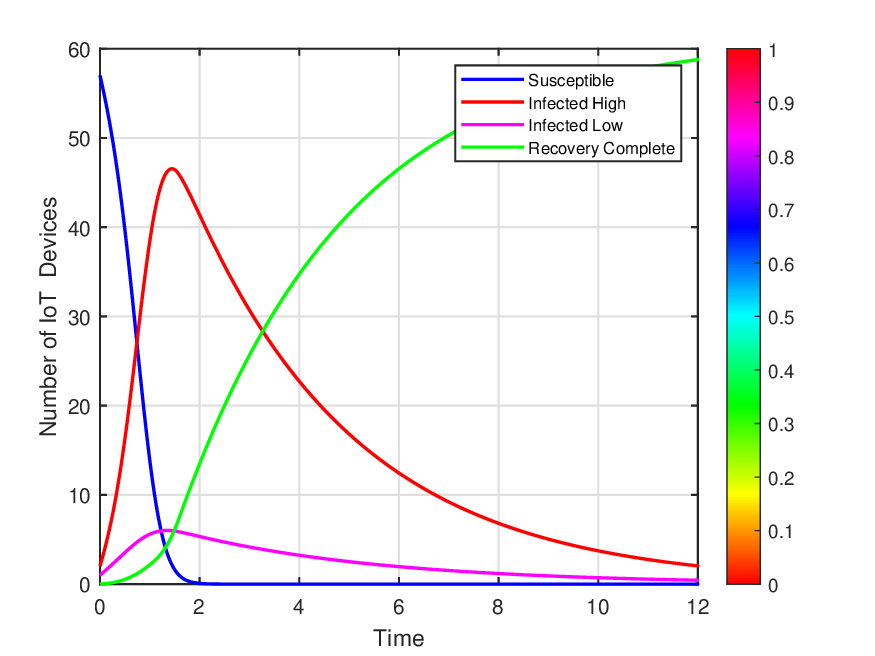}
          \vspace*{-3mm}
          \caption{The $ \mathbf{SI_HI_LR_FR_C}$ model without immediate response.} 
          \label{fig_8a}
    \end{figure}

    \begin{figure}[!h]
\centering
\includegraphics[width=3.2in , height=3.0in]{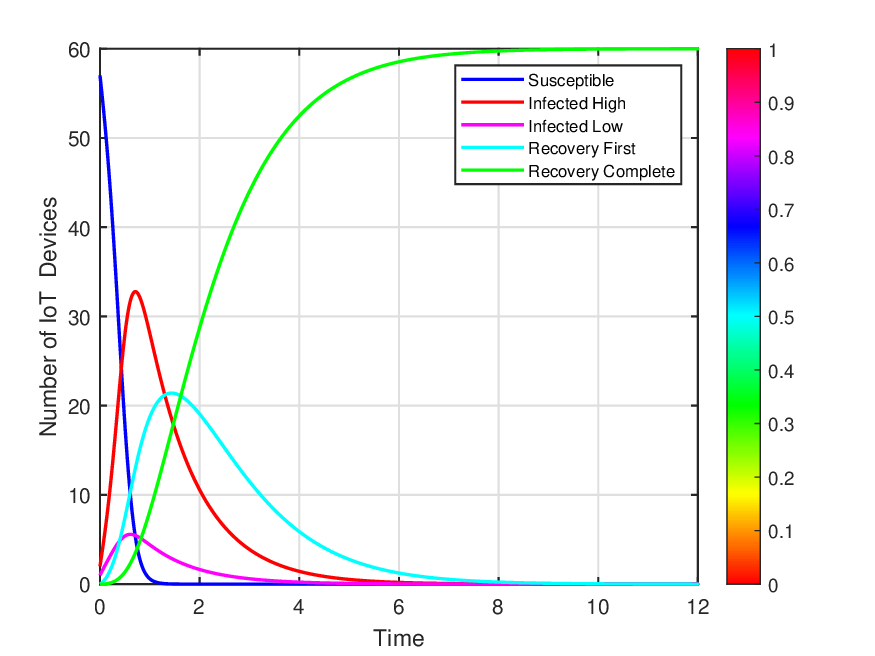}
\vspace*{-3mm}
\caption{The $ \mathbf{SI_HI_LR_FR_C}$ model with immediate response. } 
\label{fig_9a}
\end{figure}

The parameters for applying the $ \mathbf{SI_HI_LR_FR_C}$ model with restricted environment are set as the follows: \newline  
          $N=60, \; T=12 \; \beta_{H}=0.004, \; \beta_{L}=0.002, \;\gamma_{H_i}=0.4, \; \gamma_{L_i}=0.2, \; \delta_{i}=0.5.$   
          
Based on the results presented in  {\autoref{fig_9a}}, the immediate response approach has proved effective in containing the spread of malware among IoT devices with low infection rates, and has also helped to minimize the number of infected devices with high infection rates. Specifically, the number of IoT devices infected at a high rate of propagation has been successfully reduced to 33. However, it should be noted that 6 IoT devices still had infections with low propagation rates despite the implementation of the immediate response strategy.

\begin{table*} [!h]
\centering
\caption{A set of values for the parameters experiment 4.}
\centering
\begin{tabular}{m{1,8 cm} >{\centering\arraybackslash} m{.32 cm} >{\centering\arraybackslash}m{.32 cm} >{\centering\arraybackslash} m{.4 cm} >{\centering\arraybackslash} m{.4 cm} >{\centering\arraybackslash} m{.4 cm}>{\centering\arraybackslash} m{.4 cm}>{\centering\arraybackslash} m{.4cm}>{\centering\arraybackslash} m{.7cm} >{\centering\arraybackslash} m{.8cm}>{\centering\arraybackslash} m{.8cm}>{\centering\arraybackslash} m{.4cm}>{\centering\arraybackslash} m{.4cm}>{\centering\arraybackslash} m{.4cm}>{\centering\arraybackslash} m{.4cm} >{\centering\arraybackslash}m{.3cm}>{\centering\arraybackslash} m{.3cm}>{\centering\arraybackslash} m{.3cm}>{\centering\arraybackslash} m{.3cm}}\hline
parameters  & N & T & $\mathbf{S_i}$ & $\mathbf{I_{H_i}}$ & $\mathbf{I_{L_i}}$ & $\mathbf{R_{F_i}}$ & $\mathbf{R_{C_i}}$ & $\beta_{H}$ &$\beta_{L}$ &$\gamma_{H_i}$ &$\gamma_{L_i}$ &$\delta_{i}$ & $\overline{\gamma_{L_i}}$ & $\underline{\gamma_{L_i}}$ &$\overline{\gamma_{H_i}}$  &$\underline{\gamma_{H_i}} $  & $\overline{\delta_{i}}$ & $ \underline{\delta_{i}}$\\  \hline
Stage I &  60  & 30 & 57 & 2 & 1  & 0  & 0  & 0.0021 & 0.0020 & 0.35 & 0.2 & 0.6 & 0.2 & 0.1 & 0.3 & 0,1 & 0.6 & 0.1  \\ 
Stage II &  60  & 30 & 57 & 2 & 1  & 0  & 0  & 0.0022 & 0.0020 & 0.35 & 0.2 & 0.6 & 0.2 & 0.1 & 0.3 & 0,1 & 0.6 & 0.1   \\ 
Stage III &  60  & 30 & 57 & 2 & 1  & 0  & 0  & 0.0023 & 0.0020 & 0.35 & 0.2 & 0.6 & 0.2 & 0.1 & 0.3 & 0,1 & 0.6 & 0.1   \\ 
Stage IV &  60  & 30 & 57 & 2 & 1  & 0  & 0  & 0.0024 & 0.0020 & 0.35 & 0.2 & 0.6  & 0.2 & 0.1 & 0.3 & 0,1 & 0.6 & 0.1  \\    \hline
\end{tabular}
\label{tab:Table4}
\end{table*}

After conducting experiment 3, the $ \mathbf{SI_HI_LR_FR_C}$ model demonstrated that the immediate response was effective in containing infected devices with low infection rates and reducing the number of infected devices with high infection rates. The infection rate of IoT devices was reduced by 21.66\%, with the number of devices infected with high infection rates dropping from 46 devices (76.66\%) to 33 devices (55\%) after applying the restricted environment. To examine the difference between using the restricted environment and not using it, snapshots were taken at the peak of the infection process and are presented in  {\autoref{fig_10a}} and  {\autoref{fig_11a}}.

  \begin{figure}[!h]
\centering
\includegraphics[width=3.3in , height=2.4in]{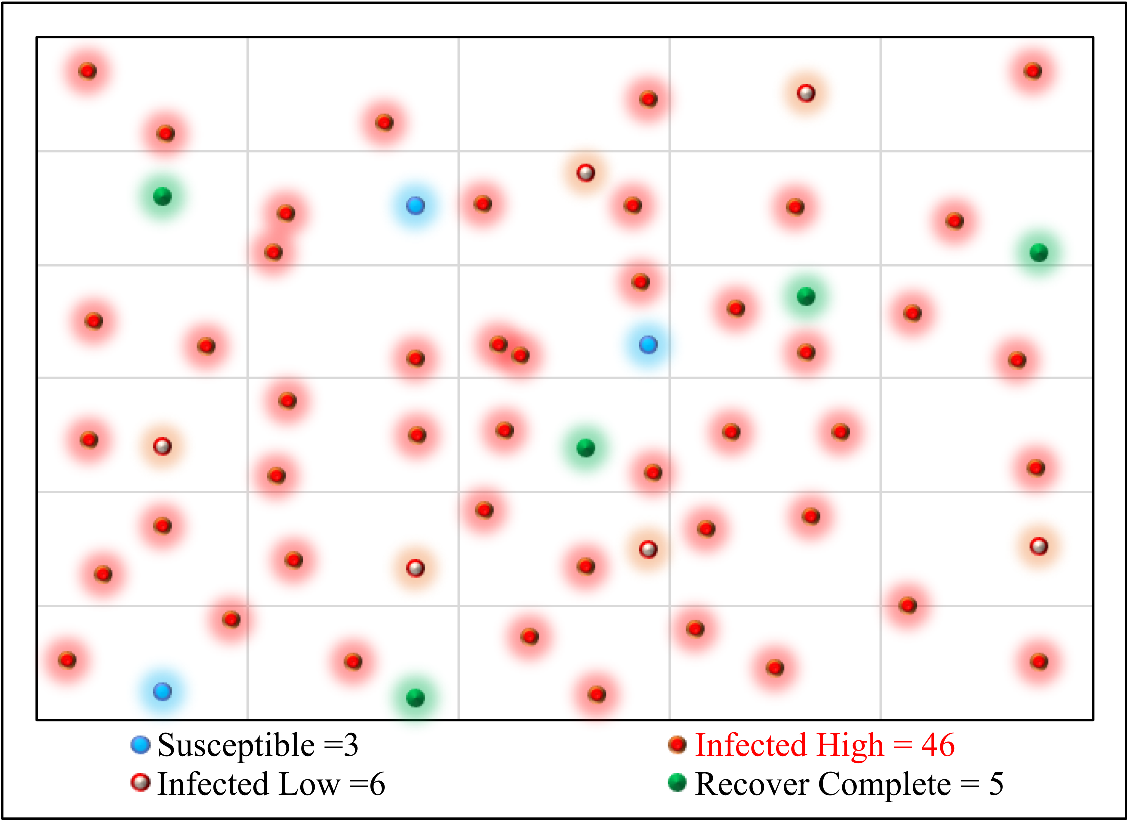}
\vspace*{-4mm}
\caption{The snapshot of the node states taken during the peak infection time without receive an immediate response.} 
\label{fig_10a}
\end{figure}

 \begin{figure}[!h]
\centering
\includegraphics[width=3.3in , height=2.4in]{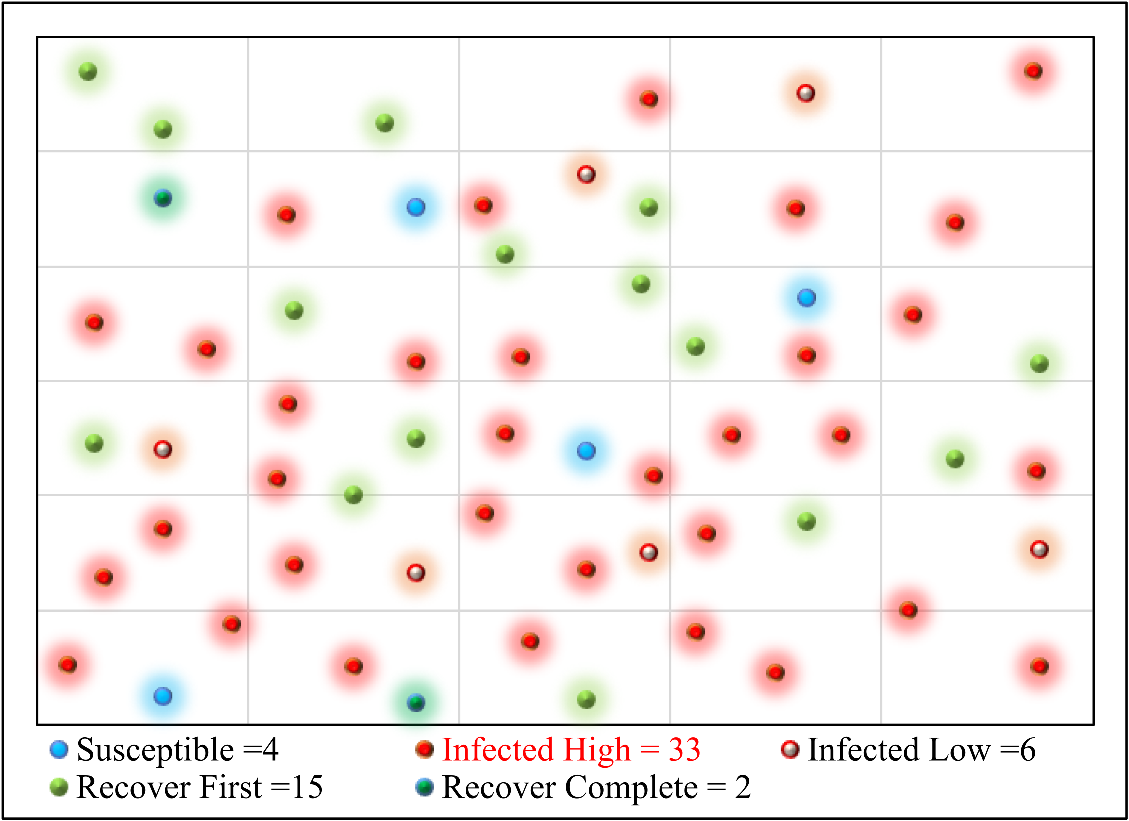}
\vspace*{-4mm}
\caption{The snapshot of the node states taken during the peak infection time with receive an immediate response.
 } 
\label{fig_11a}
\end{figure}

 \begin{figure*}[htbp] 
	    \centering
	    \subfigure[]{
		    \begin{minipage}[t]{0.45\linewidth} 
			    \centering
			      \includegraphics[width=3.3in, height=2in]{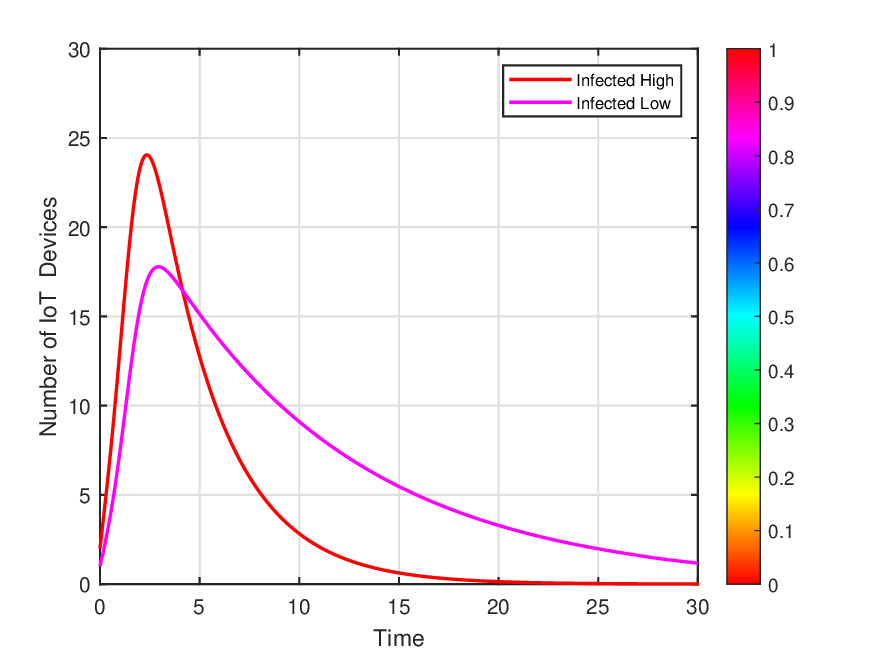} 
		      \end{minipage}
	    }
	    \hspace{0.2in}
	    \subfigure[]{
		    \begin{minipage}[t]{0.45\linewidth}
			    \centering
			    \includegraphics[width=3.3in, height=2in]{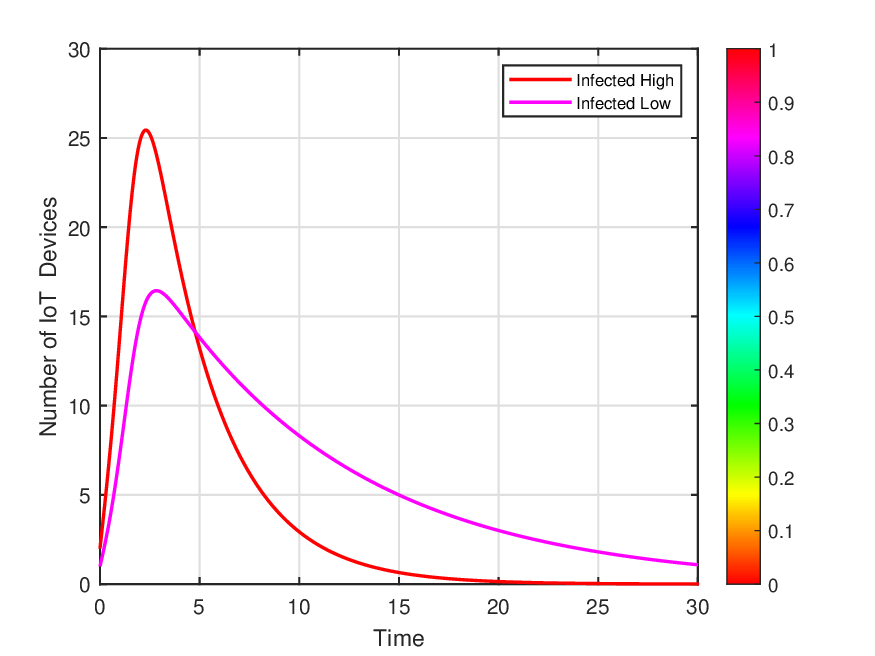}
		    \end{minipage}%
	    }
 	    \subfigure[]{
		    \begin{minipage}[t]{0.45\linewidth}
		        \centering
			    \includegraphics[width=3.3in, height=2in]{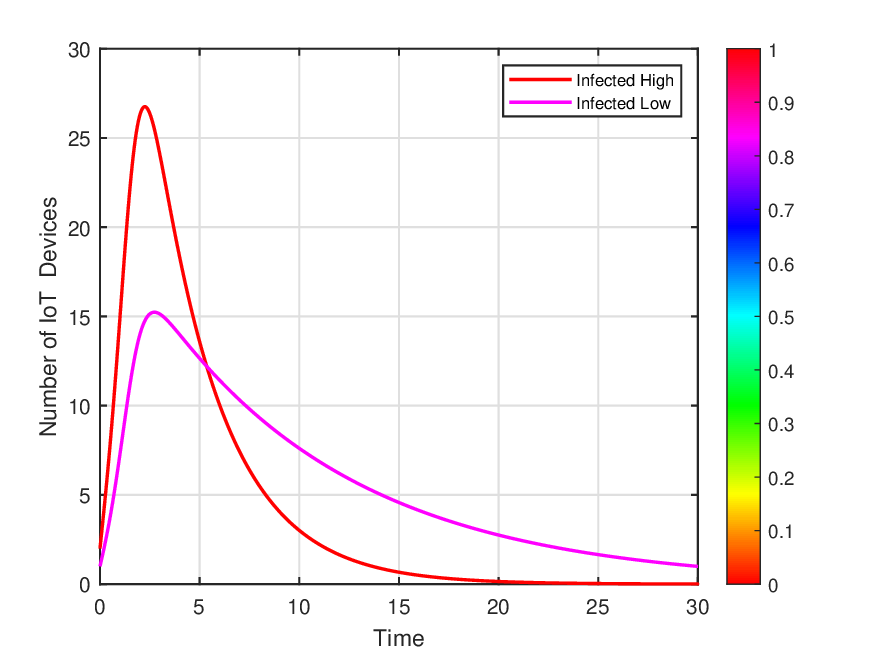}
		    \end{minipage}
	    }
	    \hspace{0.2in}
	    \subfigure[]{
		    \begin{minipage}[t]{0.45\linewidth}
		           \centering
		             \includegraphics[width=3.3in, height=2in]{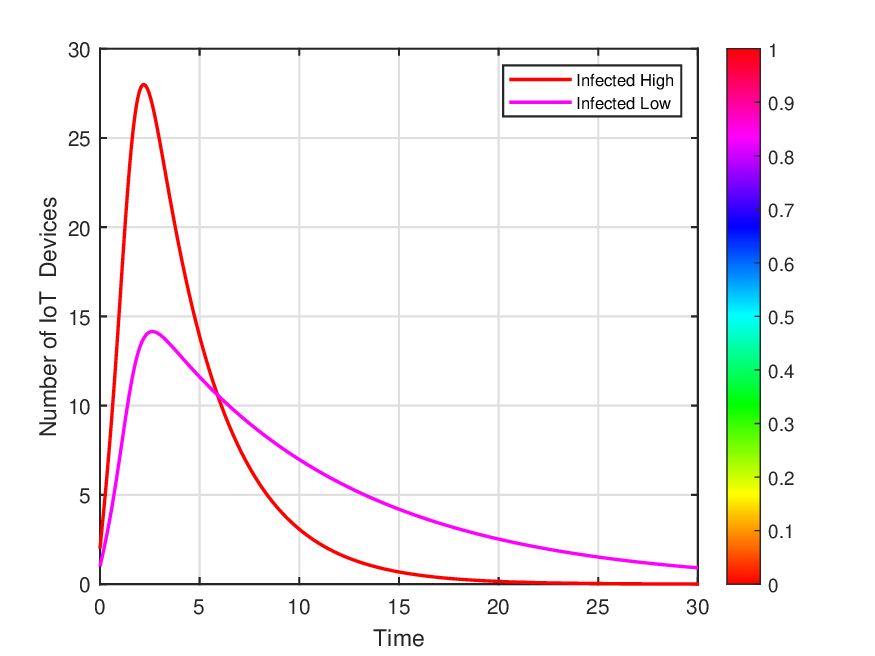}
		    \end{minipage}
	    }
         \vspace*{-3mm}
 	    \caption{The results of experiment 4 at four different stages are as follows: (a) $\beta_{H} = 0.0021 $ for stage I, (b) $\beta_{H} = 0.0022 $ for stage II, (c) $\beta_{H} = 0.0023 $ for stage III, and (d)$\beta_{H} = 0.0024 $ for stage IV.}
           \label{fig12}
	\end{figure*}

\textbf{Experiment 4:} A comparison is conducted to assess the propagation capability between malware with a high infection rate and malware with a low infection rate.

         The malware\textquotesingle s propagation potential was evaluated to differentiate between devices infected with malware that had a high propagation capability and those that had a low propagation capability within the framework of optimal control strategy.
         Experiment 4 consisted of four stages, wherein the infected high capability was gradually increased from 0.0021 to 0.0024 across the stages. Specifically, the infected high capability was 0.0021 in the first stage, 0.0022 in the second stage, 0.0023 in the third stage, and 0.0024 in the final stage. Throughout all of these stages, the infected low capability remained constant at a value of 0.0020. The values of the parameters for system 12 have been specified in \autoref{tab:Table4}.
         
As depicted in \autoref{fig12}a, during stage I, the number of IoT devices infected with high propagation capability reached 24 devices, while the number of IoT devices infected with low propagation capability was 17 devices.\newline   
While in stage II, the number of IoT devices that were infected with a high propagation capability reached  25 devices, whereas the number of IoT devices which infected with a low propagation capability was 16 devices as shown in {\autoref{fig12}}b.

Subsequently, as illustrated in {\autoref{fig12}}c, during stage III, the number of IoT devices infected with high-propagation capability was 26 devices, whereas the number of IoT devices infected with low-propagation capability was 15 devices.

Finally in stage IV, the number of IoT devices infected with high propagation capability reached 27 devices, while the number of IoT devices infected with low propagation capability was 14 devices as shown in {\autoref{fig12}}d.

During all stages, the growth rates of infected IoT devices were noticeably different between those with high and low propagation capabilities. Specifically, the number of infected devices with high propagation capability increased at a faster rate than those with low propagation capability, while the latter experienced a decrease. 

The experiment 4 results showed that the $ \mathbf{SI_HI_LR_FR_C}$ model revealed an important finding: an IoT device infected with a high propagation capability not only affects the spread of malware but also contributes to an increase in the number of infected devices.

\section{Conclusion}
\label{Conclusion}
If IoT network is targeted by a cyber-attack and breached successfully, it is possible that the new patch will not be obtainable for a prolonged duration. This paper presents a solution to the problem of mitigating cyber-attacks prior to the implementation of a new patch. Specifically, we propose an optimal control strategy that effectively curtails the spread of malware and minimizes the number of compromised devices in IoT networks. Subsequently, the optimal control strategy was executed to minimize the expenses incurred in the deployment of the latest and most potent patch. Additionally, the expenses associated with the implementation of the restricted environment and other security measures were also minimized.

We have developed a novel epidemiological model that is node-based and takes into account five distinct categories: susceptible, infected high, infected low, recover first, and recover complete $(\mathbf{SI_HI_LR_FR_C})$ . The model also incorporates an immediate response state tailored for restricted environments. The proposed model introduces a new strategy to combat the propagation of malware. By implementing this approach, a reduction of 21.66\% in the infection rate of IoT devices can be achieved. Prior to applying this technique, 76.66\% of the devices were infected with a high infection rate. However, after the implementation of the restricted environment, the infection rate dropped to 55\%.

\appendix
\section{Appendix}
\label{Appendixa}
{\appendix[Proof of the existence of an optimal control]

{\bf{Lemma 1.}} There is \; $\mathbf{u}(t) \in \mathbf{U}$   \; such that system (12) is solvable.\newline
$Proof$. \;Consider the following uncontrolled system :
\begin{equation}
\label{deqn_ex1a}
{\frac{d\mathbf{E}(t)}{dt}} = f(\mathbf{E}(t),\overline{\mathbf{u}}(t)), \qquad  0\leq t \leq T. 
\end{equation}
Where $\mathbf{u}(t) \equiv \overline{\mathbf{u}}= (\overline{\delta},...,\overline{\delta}, \; \overline{\gamma_{H}},...,\overline{\gamma_{H}}, \;\overline{\gamma_{L}},...,\overline{\gamma_{L}})^T$\; into system (12) with $\mathbf{E}(0)\in \Omega $. \newline
It is simple to check that the function $f(\mathbf{E},\overline{\mathbf{u}})$ is continuously differentiable, and $\Omega $ is positively invariant for the system. Then the claim follows from the continuation theorem for differential dynamical Systems. \newline

{\bf{Lemma 2.}} The admissible control set $\mathbf{U}$ is convex.\newline
$Proof$. \; Let \newline \newline
$\mathbf{u}^{(1)} = (\delta^{(1)}_1,...,\delta^{(1)}_N, \; \gamma_{H_1}^{(1)},...,\gamma_{H_N}^{(1)}, \; \gamma_{L_1}^{(1)},...,\gamma_{L_N}^{(1)})^T     \;   \in \mathbf{U},$ \newline 

$\mathbf{u}^{(2)} = (\delta^{(2)}_1,...,\delta^{(2)}_N, \; \gamma_{H_1}^{(2)},...,\gamma_{H_N}^{(2)}, \; \gamma_{L_1}^{(2)},...,\gamma_{L_N}^{(2)})^T \;  \in \mathbf{U},$ \newline

and $0< w < 1$. \newline

As $(L^2[0, T]^N)$ is a real vector space, then we can get \newline

$ (1-w){\bf{\mathbf{u}}}^{(1)} + w {\bf{\mathbf{u}}}^{(2)} \; \in (L^2[0, T]^N)$. \newline
 
Then, the claim follows from the  observation  that:  \newline
 
$\underline{\delta_{i}} \leq (1-w){\bf{\mathbf{u}}}^{(1)} + w {\bf{\mathbf{u}}}^{(2)}   \leq  \overline{\delta_{i}},$ \newline

$  \underline{\gamma_{H_i}} \leq (1-w){\bf{\mathbf{u}}}^{(1)} + w {\bf{\mathbf{u}}}^{(2)}   \leq   \overline{\gamma_{H_i}},$ \newline
 
$ \underline{\gamma_{L_i}} \leq (1-w){\bf{\mathbf{u}}}^{(1)} + w {\bf{\mathbf{u}}}^{(2)}   \leq  \overline{\gamma_{L_i}},$ \newline

$ 1\leq i \leq N, \; \; 0\leq t \leq T. $ \newline

{\bf{Lemma 3.}} The admissible set $\mathbf{U}$ is closed.\newline

$Proof$. \;Let $ {\bf{\mathbf{u}}}= (\delta_1,...,\delta_N, \; \gamma_{H_1},...,\gamma_{H_N}, \; \gamma_{L_1},...,\gamma_{L_N})^T$ be a limit point of  $\mathbf{U}$ and \newline

 $ {\bf{\mathbf{u}}}^{(n)}= (\delta_1^{(n)},...,\delta_N^{(n)}, \; \gamma_{H_1}^{(n)},...,\gamma_{H_N}^{(n)}, \; \gamma_{L_1}^{(n)},...,\gamma_{L_N}^{(n)})^T \newline \; \; \in U , \qquad     n=1,2,....., $ \newline

 be a sequence of points in $\mathbf{U}$ that approaches ${\bf{\mathbf{u}}}$.  \newline

$\parallel {\bf{\mathbf{u}}}^{(n)} - {\bf{\mathbf{u}}} \parallel _2 \; :=    \left[ \int_{0}^{T} \; \;| {\bf{\mathbf{u}}}^{(n)}(t) - {\bf{\mathbf{u}}}(t) |^2  \; \;  dt \right]^{\frac1{2}}     < \frac1{n},$ \newline

As $(L^2[0, T]^N)$ is complete, then we can get: \newline

\[ \lim_{n \to \infty} {\bf{\mathbf{u}}}^{(n)} =  {\bf{\mathbf{u}}}  \in (L^2[0, T])^N ,\]    \newline

Thus, the closeness of $\mathbf{U}$ and the claim follows from the observation that  \newline
\[  \underline{\delta_{i}} \leq  \delta_i(t)= \lim_{n \to \infty} \delta_i^{(n)}(t)  \leq  \overline{\delta_{i}},  \]\newline
\[   \underline{\gamma_{H_i}} \leq  \gamma_{H_i}(t)= \lim_{n \to \infty} \gamma_{H_i}^{(n)}(t)   \leq   \overline{\gamma_{H_i}},\]  \newline
 \[  \underline{\gamma_{L_i}} \leq  \gamma_{L_i}(t)= \lim_{n \to \infty} \gamma_{L_i}^{(n)}(t)   \leq  \overline{\gamma_{L_i}}, \] 
\begin{flushright}$ 1\leq i \leq N, \; \; 0\leq t \leq T. $ \end{flushright} 

{\bf{Lemma 4.}} $ {\bf{f(\mathbf{E}, \mathbf{u})}}$ is bounded by a linear function in $\bf{\mathbf{E}}$.\newline
$Proof$. \;Note that, the claim follows from the observation, for the following system and  $ \; \;0\leq t \leq T ,\; \; i=1,2,....,N,  $  

\begin{equation*}
          \begin{split}
             & {\frac{dI_{H_i}(t)}{dt}} = \beta_{H} \; S_i(t)\; \sum_{j=1}^{N} a_{ij} \;  I_{H_j}(t) -\gamma_{H_i}(t)I_{H_i}(t) , \\ 
             & \qquad \qquad \qquad \qquad \qquad  \qquad  \qquad t \geq 0, \; 1\leq i \leq N,\\
             & {\frac{dR_{F_i}(t)}{dt}} =  \gamma_{H_i}(t)I_{H_i}(t) + \gamma_{L_i}(t)I_L(t) - \delta_{i}(t)R_{F_i}(t),  \qquad \qquad, \\
             & \qquad \qquad \qquad \qquad \qquad  \qquad  \qquad t \geq 0, \; 1\leq i \leq N,\\
             & {\frac{dR_{C_i}(t)}{dt}} = \delta_{i}(t)R_{F_i}(t), \qquad \qquad  t \geq 0, \;  1\leq i \leq N,\\
            \end{split}
    \end{equation*}

Then we can get: 
\begin{equation*}
          \begin{split}
               & - \; \overline{\gamma_{H_i}} \; I_{H_i}  \leq \beta_{H} \; S_i(t) \; \sum_{j=1}^{N} a_{ij} \;  I_{H_j} -\gamma_{H_i}I_{H_i}  \\
               & \qquad \qquad  \; \leq \beta_{H} \; S_i(t) \;\sum_{j=1}^{N} a_{ij} \;  I_{H_j} - \; \underline{\gamma_{H_i}} \; I_{H_i},\\
               & \underline{\gamma_{H_i}} \; I_{H_i} +  \underline{\gamma_{L_i}} \; I_{L_i} - \overline{\delta_{i}} \; R_{F_i} \leq \gamma_{H_i}\; I_{H_i} + \gamma_{L_i} \;I_{L_i} - \delta_{i} \;R_{F_i} \\
               & \qquad \qquad \qquad \qquad \qquad \leq  \overline{\gamma_{H_i}} \; I_{H_i} +  \overline{\gamma_{L_i}} \; I_{L_i} - \underline{\delta_{i}} \; R_{F_i},\\
               & \underline{\delta_{i}} \; R_{F_i} \leq \delta_{i} \; R_{F_i} \leq \overline{\delta_{i}}  \; R_{F_i},\\
            \end{split}
    \end{equation*}

{\bf{Lemma 5.}} $L(\mathbf{E},\mathbf{u})$ is convex on $\mathbf{U}$. \newline \newline
$Proof$. \;Let \newline \newline
$\mathbf{u}^{(1)} = (\delta^{(1)}_1,...,\delta^{(1)}_N, \; \gamma_{H_1}^{(1)},...,\gamma_{H_N}^{(1)}, \; \gamma_{L_1}^{(1)},...,\gamma_{L_N}^{(1)})^T   \; \;  \in \mathbf{U},$ \newline \newline
$\mathbf{u}^{(2)} = (\delta^{(2)}_1,...,\delta^{(2)}_N, \; \gamma_{H_1}^{(2)},...,\gamma_{H_N}^{(2)}, \; \gamma_{L_1}^{(2)},...,\gamma_{L_N}^{(2)})^T  \; \;  \in \mathbf{U},$ \newline

and let  $0< w < 1$.  Then,  
\begin{equation*}
        \begin{split}
                & L(\mathbf{E}, (1-w){\bf{\mathbf{u}}}^{(1)}(t) + w {\bf{\mathbf{u}}}^{(2)}(t))= \\
                & \sum_{i=1}^{N} \left[(I_{H_i}(t)) + \frac{1}{2} \left[\delta_{i}^{(1)}(t) + \delta_{i}^{(2)}(t) \right]^{2}  \right] \\
                & +  \sum_{i=1}^{N} \left[ \frac{1}{2}  \left[ \left[ \gamma_{H_i}^{(1)}(t) +  \gamma_{H_i}^{(2)}(t) \right]^{2} + \left[ \gamma_{L_i}^{(1)}(t) +  \gamma_{L_i}^{(2)}(t)\right]^{2} \right] \right] \\
                & \leq \sum_{i=1}^{N} I_{H_i}(t) + \sum_{i=1}^{N} \frac{1}{2} \left[ \left[\delta_{i}^{(1)}(t) \right]^{2} + \left[\delta_{i}^{(2)}(t) \right]^{2} \right]  \\ 
                & + \sum_{i=1}^{N} \frac{1}{2} \left[ \left[  \left[ \gamma_{H_i}^{(1)}(t)\right]^{2} + \left[\gamma_{H_i}^{(2)}(t) \right]^{2} \right] + \left[ \left[\gamma_{L_i}^{(1)}(t)\right]^{2} +  \left[\gamma_{L_i}^{(2)}(t)\right]^{2} \right] \right] \\
                 & \leq  \; L(\mathbf{E}, {\bf{\mathbf{u}}}^{(1)}(t)) + \; L(\mathbf{E},{\bf{\mathbf{u}}}^{(2)}(t)),  \\       
        \end{split}
    \end{equation*}

This will make the proof completed.  \newline

{\bf{Lemma 6.}} \; $L(\mathbf{E},\mathbf{u}) \geq  c_1 \parallel {\bf{\mathbf{u}}} \parallel_2 ^{\rho}  + c_2 \; for \; some \; vector\;\newline norm  \parallel \bullet \parallel, {\rho} >1 , \; c_1> 0 \; and \; c_2 . $ \newline

$Proof$. \;Let $\rho =2 , \; c_1= \frac {min_i}{2} , \; c_2=0 , $ \newline

Then  \newline

$L(\mathbf{E},\mathbf{u}) \geq  \frac {min_i}{2} E \parallel {\bf{\mathbf{u}}} \parallel_2^2 \; ,$ \newline

This will make the proof completed.\newline

\bibliographystyle{cas-model2-names}

\bibliography{cas-refs}


\end{document}